\documentclass{article}
\usepackage{graphicx} 

\graphicspath{ {./images/} }
\usepackage{url}
\usepackage{amsmath}
\usepackage{amsfonts}
\usepackage{booktabs}
\usepackage{tabularx}
\usepackage{caption}

\usepackage[
left=1in,
right=1in,
top=1in,
bottom=1.5in
]{geometry}

\usepackage{cleveref}
\usepackage{subcaption} 
\usepackage{listings}
\usepackage{xcolor} 
\usepackage{authblk}

\usepackage[textwidth=3cm,textsize=scriptsize]{todonotes}

\title{A City-Scale Dataset of Traffic Flows, Travel Times, and Urban Context}
\author[1]{Riccardo Cappi \thanks{Corresponding author: riccardo.cappi@phd.unipd.it}}
\author[2]{Massimiliano Luca}
\author[3]{Pietro Fontolan}
\author[1]{Nicolò Navarin}
\author[2]{Bruno Lepri}
\author[1,2]{Alessandro Sperduti}

\affil[1]{University of Padua, Department of Mathematics, Padua, Italy}
\affil[2]{Fondazione Bruno Kessler, Trento, Italy}
\affil[3]{Innovation and Digital Transition Sector, Municipality of Padua, Padua, Italy}
\date{September 2026}

\begin{document}

\maketitle

\begin{abstract}
We present a multi-source traffic dataset derived from Automatic Vehicle Identification (AVI) recordings in Padua, Italy, spanning from February 2026 to \textcolor{black}{August 2026}. The dataset combines traffic volume time series, aggregated at 10-minute intervals, with time-varying trajectory-based flow statistics including transition probability matrices, average travel times, and flow residuals. To enrich the traffic measurements with urban contextual information, we integrate Points Of Interest (POIs), demographic data, meteorological variables, and road infrastructure data. All components are accessible through a Python class that loads temporal and contextual data exploiting a spatio-temporal graph representation. Validation analyses confirm that the dataset captures expected traffic patterns, such as morning and evening rush hours, as well as weekdays vs. weekend days traffic routines. 
\end{abstract}

\section{Background \& Summary}
{\color{black} Traffic, pollution, disease spreading, are just some of the many societal challenges that urban policymakers are facing \cite{luca2021survey,barbosa2018human}. More broadly, the quantitative study of human dynamics has proven essential for understanding and managing a wide array of complex societal phenomena. In the realm of public health, high-resolution mobility and contact networks dictate the spatial and temporal diffusion of viral diseases, providing the foundational data necessary for epidemic forecasting and targeted containment strategies \cite{eubank2004modelling, colizza2007modeling, perkins2014theory}. During acute crises, analyzing collective behavioral responses to natural disasters reveals how populations alter their movement and communication patterns under extreme stress, offering critical insights into emergency management and societal resilience \cite{bohorquez2009common, bagrow2011collective}. Within the context of daily urban operations, mapping human trajectories is fundamental to the optimization of traffic volumes and the development of intelligent transportation infrastructures that alleviate systemic congestion \cite{batty2013new, mazzoli2019field}. Furthermore, the ways in which individuals move and interact within urban environments directly underpin macroscopic socio-economic outcomes; dense, well-connected human dynamics consistently scale with accelerated economic growth, higher rates of technological innovation, and deeper social integration \cite{bettencourt2007growth, pan2013urban, schlapfer2014scaling}. Conversely, these same underlying behavioral patterns and scaling laws are intrinsically linked to the environmental footprint of cities, directly dictating the severity of air pollution and driving the collective consumption of energy, water, and other vital physical resources \cite{bettencourt2007growth, bettencourt2010unified}.
}
Luckily, we are in the middle of a sensing revolution. Sensing technologies are becoming cheaper and more ubiquitous than ever before, giving the possibility to have access to many unconventional data streams \cite{luca2021survey, barbosa2018human}.  Automatic License Plate Recognition (ALPR) systems are not an exception and are becoming increasingly popular in modern smart cities. These systems enable the identification of vehicles at multiple locations over the road network by tracking the hashed code of license plates, making it possible to reconstruct partial trajectories and derive more fine-grained traffic flow information \cite{qi2021vehicle, rao2018origin, yu2018vehicle, mo2020estimating, liu2025vehicle}. This represents a major advancement over the past generation of traffic-monitoring sensors, such as loop detectors, which provide less information with respect to ALPR and can only measure traffic volumes at fixed locations \cite{spanu2021smart, hadavi2020analyzing, he2023link}. 
Compared with other data sources commonly used to measure traffic and trajectories, such as mobile phone or GPS data \cite{luca2021survey, luca2023crime}, ALPR-based data may have lower spatial resolution depending on the number and distribution of sensors installed. However, it offers an important advantage: it can provide potentially comprehensive coverage of vehicle mobility within the monitored road network that GPS data providers and mobile phone data providers cannot offer \cite{luca2021survey, luca2023crime}. In this sense, ALPR data are closer to traditional traffic surveys, while avoiding some of their main limitations, as surveys are typically costly, time-consuming, and difficult to update in a timely manner \cite{luca2021survey, luca2023crime}. 
\\\\
Despite the opportunities provided by modern ALPR systems, the majority of open datasets based on traffic sensors on private vehicle traffic are based on fixed-point detection technologies, without tracking vehicle routes throughout the street network \cite{DBLP:conf/iclr/LiYS018, li2024high, li2025high}. Other datasets collected via check-in systems, GPS signals, or mobile phone data are commonly limited to aggregated origin-destination (OD) matrices and, in any case, limited to specific vehicle types
\cite{chai2018bike, zhang2021traffic, beneduce2025pyspainmobility}.
For example, multiple US cities provide OD matrices for their bike-sharing systems \cite{chai2018bike, zhang2021traffic, wang2021spatio}. In other cases, OD matrices of taxi flows are provided \cite{DBLP:journals/tkde/LiZDXCLLZHP23, DBLP:journals/tkde/SunZLYLZ22, liu2022msdr, ye2021coupled}. Concerning mobile phone data, the number of open datasets is limited and provided in the form of OD matrices. However, in such cases, the datasets do not necessarily regard vehicular movements, and thus, it becomes challenging to analyze traffic dynamics \cite{barlacchi2015multi, blanchard2025highly, bergroth202224, kang2020multiscale, pappalardo2023dataset, yabe2024yjmob100k, flores2024mobility}.   
In addition, when trajectories are provided instead of OD matrices, the number of individuals is limited and generalized city-scale conclusions are hard \cite{krataithong2022taxi, larroya2023home, yuan2010t}.
Concerning traffic sensors, recently, a few datasets have been created by extracting vehicle trajectories from Automatic Vehicle Identification (AVI) recordings at the city-level \cite{wang2023city, ma2026city}. These datasets are derived from Xuancheng, China, in which the high presence of AVI systems throughout the city is exploited to reconstruct one-month trajectories for approximately 80,000 daily vehicles, or to build a simulation of the traffic flow for traffic control scenarios. Despite the valuable contributions, two main limitations remain. First of all, these datasets face constraints on data openness and privacy issues, as releasing raw ALPR data over prolonged periods, even if encrypted, may raise legitimate concerns regarding citizen privacy.
The second limitation, common to most mobility datasets, is the absence of multi-source contextual information representing urban and environmental features. In recent years, data-driven approaches to urban analysis have increasingly integrated multiple data sources to capture different aspects of urban conditions \cite{simini2021deep,li2025cross,zhang2023incorporating}. In this context, datasets that combine traffic measurements with contextual information provide a richer representation of urban dynamics and enable more comprehensive analyses of mobility phenomena.
\\\\
To address these limitations, we present a dataset that combines traffic measurements with rich contextual information, derived from historical recordings by AVI devices in the Italian city of Padua, a mid-sized city located in northeastern Italy (Veneto region) with roughly 210,000 inhabitants over an area of about 93 km$^2$. 
We collected approximately 400,000 vehicle passages per day from 2026-02-06 to \textcolor{black}{2026-08-06}, capturing distinct traffic patterns on weekdays and weekends, as well as during the mid-February school closure period. \textcolor{black}{We processed ALPR raw data to obtain time series of traffic volumes at sensor locations, aggregated in 10-minute intervals. Additionally, we collected individual vehicle paths and derived time-varying aggregated flow statistics that describe average traffic patterns across the road network, without including sensitive individual vehicle trajectories. These statistics include (i) the transition probability matrix, representing the probability of observing vehicle passages between each pair of sensors, (ii) the average travel times between sensors, and (iii) the flow residuals for each sensor.} We enriched these traffic time series with contextual information derived from various sources, including Points of Interest (POIs), demographic data aggregated by census section, meteorological variables, and the city road network. Finally, we release a ready-to-use Python code to organize traffic volume time series into sliding windows, in which flow-based statistics and contextual information are dynamically linked to AVI sensors through a graph-based representation.
\section{Methods}
In this section, we first detail the data acquisition process of both temporal and contextual data. Then, we describe the pre-processing steps performed to clean the data and we formally define the derived trajectory-based flow statistics. Finally, we present the feature enrichment process used to build a graph-based representation of AVI sensors integrating contextual information and traffic flow data.  
\subsection{Data Acquisition}
\begin{figure}[h]
    \centering
    \begin{subfigure}[b]{0.48\linewidth}
        \centering
        \includegraphics[width=\linewidth]{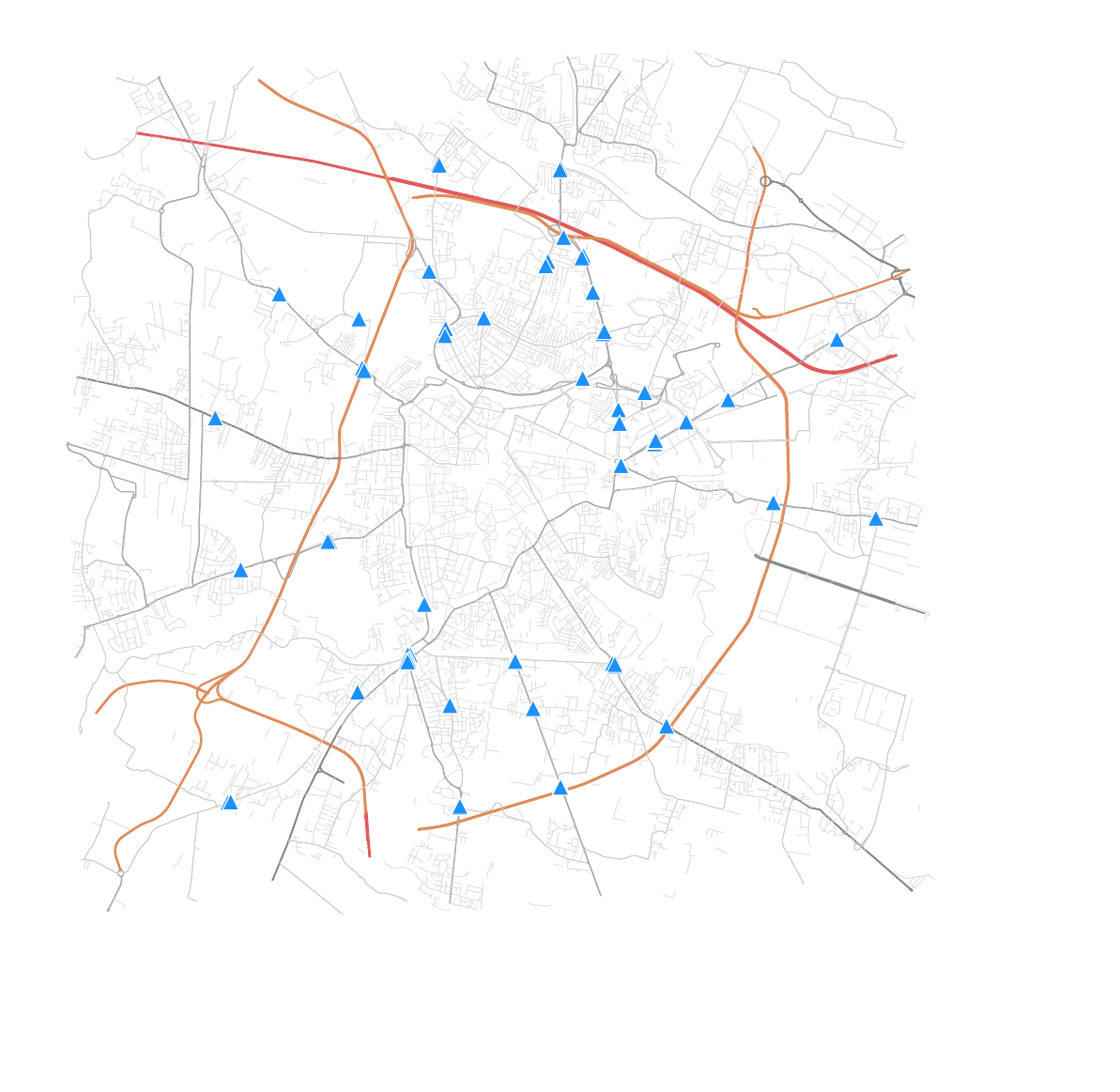}
        \caption{}
        \label{fig:sensors_a}
    \end{subfigure}
    \hfill
    \begin{subfigure}[b]{0.48\linewidth}
        \centering
        \includegraphics[width=\linewidth]{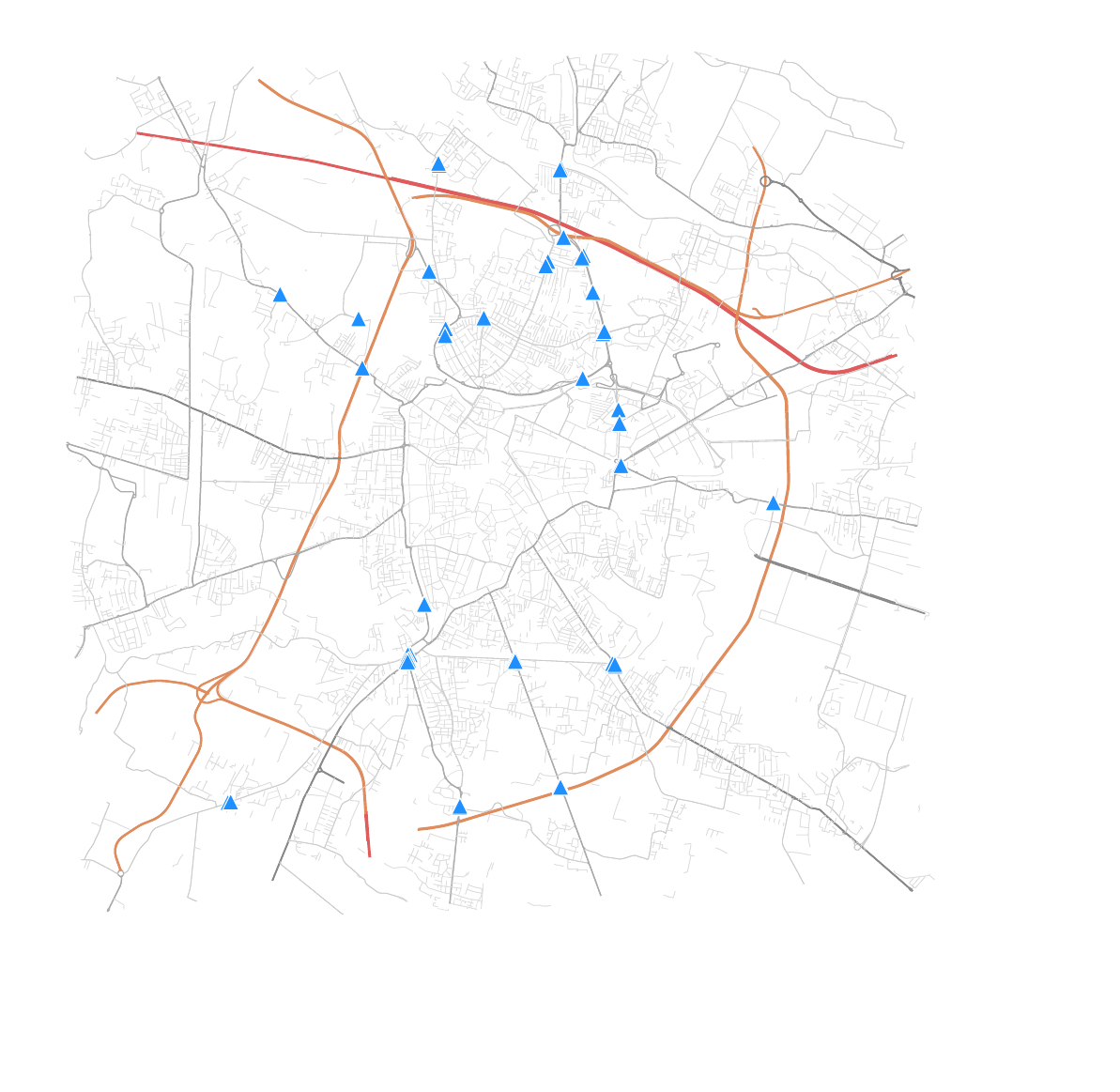}
        \caption{}
        \label{fig:sensors_b}
    \end{subfigure}
    \caption{(a) Spatial distribution of all the 96 AVI sensors, and (b) the 70 remaining devices after \textcolor{black}{removing inactive sensors}. Blue triangles indicate sensor locations. Note that the collocation of multiple sensors on different arms or approaches of the same intersection causes markers to overlap at this scale.}
    \label{fig:sensors_spatial_distr}
\end{figure}

\paragraph{Time series data:} Historical traffic measurements were collected from a set of AVI sensors that record individual vehicle passages, together with metadata such as travel direction, timestamp, and license plate. We downloaded the raw AVI measurements from the server of the local police department after removing all personally identifiable information. In addition, license plates were encrypted using a hashing mechanism before downloading. Data access was granted under a formal collaboration agreement between the Municipality of Padua and the local police department, while the construction and release of this aggregated dataset was done in accordance with the Municipality of Padua. Data were collected from 2026-02-06 to \textcolor{black}{2026-08-06} for the 96 AVI sensors shown in Figure \ref{fig:sensors_a}, mostly located in the urban area surrounding the city center. The daylight saving time (DST) transition on 2026-03-29 was automatically handled by the sensors’ internal clocks; therefore, in the processed files, the time index between 02:00 and 03:00 on that date is not present. 
\paragraph{Contextual data:} We acquired contextual data from the following sources:
\begin{itemize}
    \item \textbf{Points of Interest (POIs)}: \textcolor{black}{Downloaded from the Overture Maps platform \cite{overture2026places}.} They provide relevant information on the spatial distribution of traffic attractors, such as shopping centers, schools, and offices. \textcolor{black}{The released code (see Section \ref{sec:code}) contains the original query used to download raw data.}

    \item \textbf{Demographic Data}: We collected the number of inhabitants per census section, provided directly by \textcolor{black}{the Innovation and Digital Transition Sector of the Municipality of Padua \footnote{\url{itd@comune.padova.it}}.} This information reflects the population density, which is correlated with potential origins and destinations of trips.

    \item \textbf{Meteorological Variables}: Data on temperature, humidity and precipitation, obtained from \textcolor{black}{ARPAV open data \cite{arpav2026meteo}} and recorded by a single station within the city. Data on city weather conditions are well known to be correlated with variations in driving behavior and traffic conditions.

    \item \textbf{Road network}: The city’s road network topology was retrieved  \textcolor{black}{from OpenStreetMap data through the \texttt{osmnx} library \cite{osm2026padua, DBLP:journals/corr/abs-2505-00736} (See Section \ref{sec:code} for code reference).} This dataset includes essential attributes such as road classification and segment lengths. Characterizing a region’s transportation capacity is essential for interpreting traffic patterns, as areas dominated by trunk roads tend to attract greater vehicle volumes due to increased infrastructure capacity \cite{li2025cross}.
\end{itemize}
Contextual data, together with the historical traffic time series and the flow-based statistics derived from ALPR, constitute the comprehensive multi-source dataset presented in this work.

\subsection{Time series pre-processing}
\label{sec:ts-pre-proc}
In this section, we present the pre-processing steps applied to the time series data, including missing data handling, temporal aggregation, and spatial clustering.

\begin{figure}[h]
    \centering
    \includegraphics[width=0.6\linewidth]{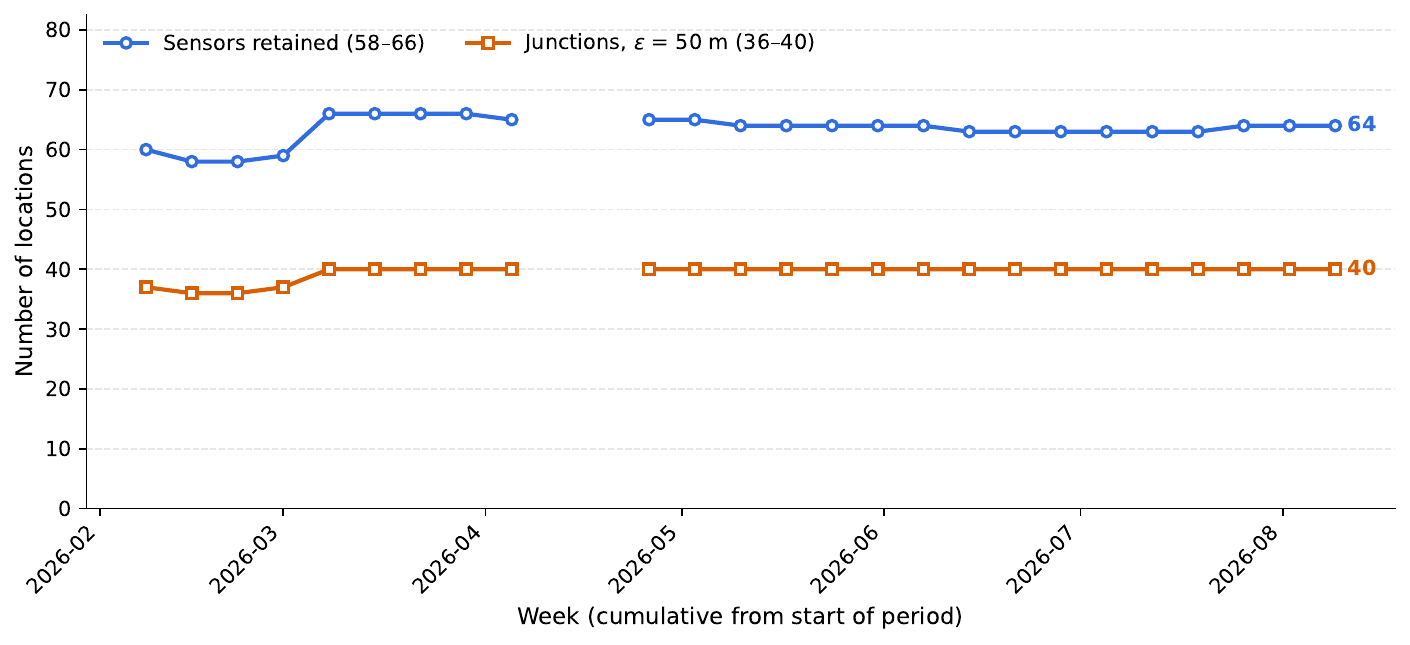}
    \caption{\textcolor{black}{Number of AVI sensors retained by the reliability filter (blue) and number of distinct junction-level locations obtained by spatially aggregating them within a
    50 m radius (orange), computed incrementally week by week. At each week the binary mask is recomputed over all observations collected up to that point, using a threshold of one hour of consecutive zero values, and excluding sensors with more than 50\% of invalid observations.}}
    \label{fig:weekly_sensor_coverage}
\end{figure}

\paragraph{Temporal aggregation \& Missing data handling:} To compute traffic volumes at each sensor location, we aggregated individual vehicle passages into consecutive 10-minute time bins by summing all detections occurring within each interval over the entire observation period. After aggregation, missing values were imputed with zeros. Missing values correspond to time windows in which no vehicle passage was recorded by a sensor. Such zeros may reflect genuinely low traffic demand (e.g., during nighttime periods) or temporary data unavailability. \textcolor{black}{First, we excluded sensors for which no vehicle passage was ever recorded, resulting in a set of 70 AVI devices shown in Figure \ref{fig:sensors_b}. The relatively large number of excluded sensors is due to infrastructure works carried out in the city during the observation period, which may have disrupted the operation of several cameras. Then, to improve the reliability of the final dataset, we identified sensors exhibiting prolonged periods of inactivity. Specifically, we computed a binary mask by examining each sensor independently and identifying contiguous segments of zero values. When the length of such a segment exceeded a predefined threshold, the corresponding observations were flagged as uninformative in the mask, while all remaining observations were retained. This procedure preserves isolated zeros that are plausibly due to low traffic, while identifying longer zero-valued sequences that are more likely indicative of sensor malfunction. Since we release time series data for all the 70 sensors, users can compute the binary mask using custom threshold for consecutive zero values and exclude any device exceeding a chosen percentage of invalid observations. For example, in Figure \ref{fig:weekly_sensor_coverage} we show the number of retained sensors over time by applying this procedure with a threshold of 1 hour for consecutive zero values, aligning the dataset with the predictive horizon considered in most existing traffic forecasting studies \cite{DBLP:conf/iclr/LiYS018, DBLP:conf/ijcai/WuPLJZ19, DBLP:conf/ijcai/YuYZ18, DBLP:journals/tkdd/LiFYJYSJL23}, and excluding any sensor for which more than 50\% of the observations were flagged as invalid. The procedure is applied incrementally week by week starting from the beginning of the observed period. The Figure shows that the number of retained sensors is not stationary: it rises to 66 as more observations accumulate and then declines to 64 by the end of the period. The number of junction-level locations (see the spatial clustering paragraph), by contrast, stabilises at 40 from the beginning of March and remains constant thereafter, showing that the spatial coverage of the network is preserved.}
\textcolor{black}{Supplementary Section \ref{app:missingness_stats} complements this analysis by quantifying the sensitivity of the mask to the consecutive-zero threshold and the distribution of invalid observations across devices.}
\\\\
\textcolor{black}{Concerning the temporal coverage, the observed period spans from 2026-02-06 to 2026-08-06. However, there are three main periods where observations were lost due to technical failures of the data acquisition server, i.e., from 2026-03-19 to 2026-03-27, from 2026-04-02 to 2026-04-26, and from 2026-05-29 to 2026-06-01. Other minor missing periods include the single day of 2026-06-09, and 2026-07-29 from 00:00 to 09:00. After excluding these gaps, each sensor yields 20,808 valid 10-minute traffic-volume records, compared with the 26,196 bins of the complete 10-minute time index.} 

\paragraph{Spatial clustering:} 
To support analyses at the junction level, we aggregated sensor time series into spatial clusters representing nearby installations. We first applied the DBSCAN algorithm \cite{DBLP:conf/kdd/EsterKSX96} to group sensors that are geographically close, using a maximum neighborhood radius of 50 meters and the Euclidean distance as metric. We then refined each spatial group by separating sensors according to the traffic flow direction they observe (see Section \ref{sec:data_records} for details on the directions observed by the sensors). This step allows geographically close sensors monitoring different directional flows to be aggregated separately. The final junction positions were computed as the centroids of the sensors within each cluster. The clustering process resulted in a set of 40 junction clusters. The choice of the neighborhood radius has a limited impact on the results: using a radius of 25 or 100 meters yields 44 and 38 junctions, respectively, suggesting that the sensor layout is relatively robust to this parameter. We selected 50 meters as an intermediate value. Since we release both the data and the code, users can readily perform the spatial aggregation with a custom radius to suit their specific needs.

\subsection{Contextual information}
\label{sec:context}
\paragraph{Points of Interest (POIs):} 
We retrieved a dataset of Padua's POIs using the API provided by Overture Maps \cite{overture2026places}.
\begin{figure}[h]
    \centering
    \includegraphics[width=0.7\linewidth]{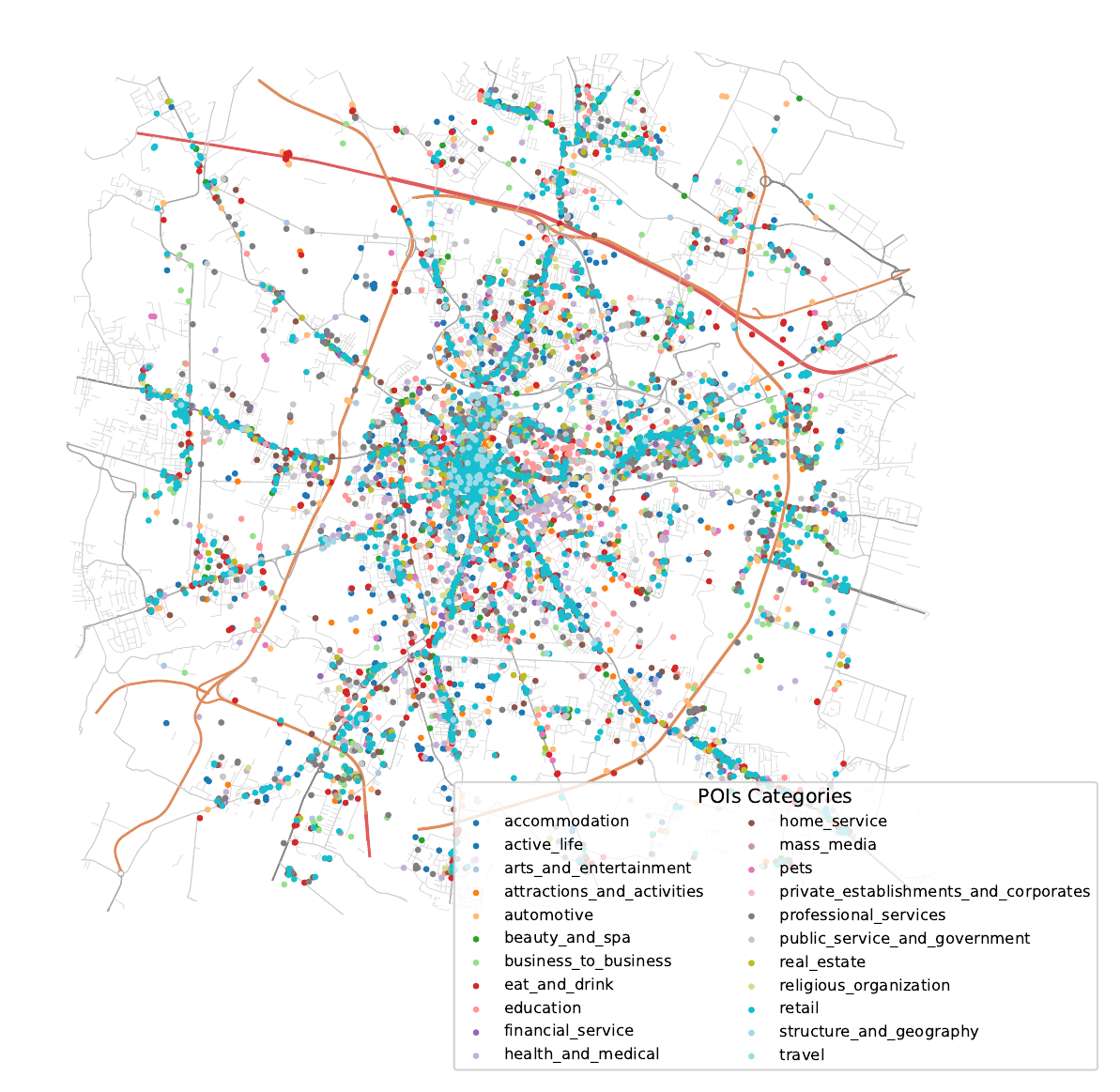}
    \caption{POIs distribution over the city, categorized in the 22 parent classes. The map shows that the majority of POIs are clustered in the city center, gradually dispersing towards the urban area.}
    \label{fig:pois}
\end{figure}
The initial POIs dataset contains 13,034 entries divided into 726 distinct categories (e.g., “pizzeria,” “bar,” “beauty salon,” etc.). Each point is associated with a score ranging from 0 to 1, representing the probability of the place’s existence. To ensure data quality, we excluded entries with a confidence parameter below 0.7, resulting in a final set of 9,262 POIs. \textcolor{black}{Note that this filtering process was performed to release a ready-to-use dataset containing POIs information with high existence probability. However, since we release both raw data and the code implementing the filter, users can adopt custom thresholds and recompute the dataset according to their needs. We refer to Supplementary Section \ref{app:pois} for the  analyses on the effect of the confidence parameter on the number of excluded POIs.}
\\\\
To reduce the dimensionality of the original category space, we implemented a hierarchical classification process. Using the class taxonomy provided by the Overture Maps platform, each POI was reclassified by assigning it to the category corresponding to the parent node closest to the root of the taxonomic tree. This approach allowed us to aggregate the 726 original categories into a manageable set of 22 distinct classes. For example, the 726 initial categories include very specific and granular classes such as “Indian restaurant” or “Thai restaurant.” Through the hierarchical classification process, these highly specific groups are aggregated under a more general parent class, such as “eat\_and\_drink.” Figure \ref{fig:pois} illustrates the processed POIs categorized into the 22 aggregated classes. \textcolor{black}{As for the confidence parameter, users can leverage the released code to reclassify each POI according to a specific level of the taxonomy tree.}

\begin{figure}[h]
    \centering
    \includegraphics[width=0.7\linewidth]{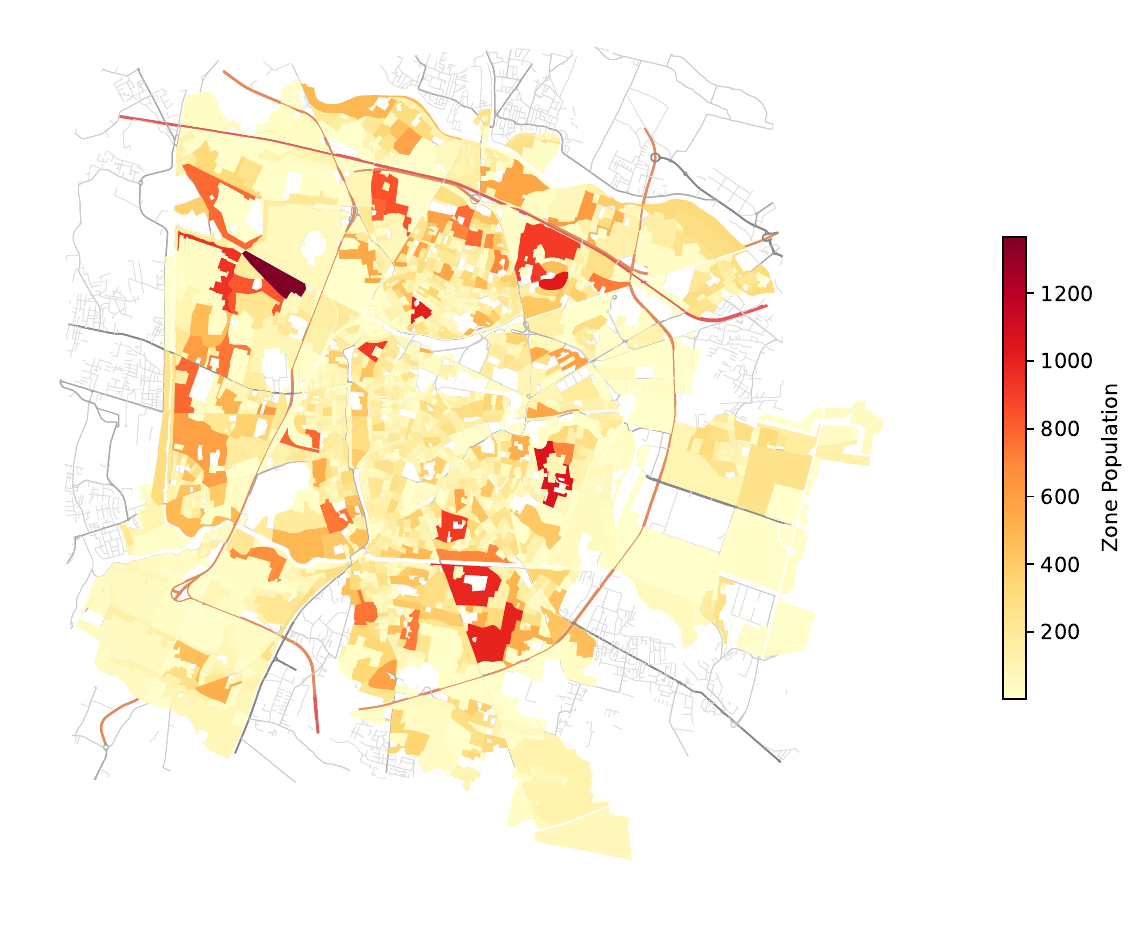}
    \caption{Spatial distribution of residential population across census zones in the city. The map colors encode zone-level population counts, ranging from yellow (sparsely populated) to red (highly populated). }
    \label{fig:zones}
\end{figure}

\paragraph{Demographic Data:} We divided the city in 1777 zones corresponding to official census sections provided by the Padua Municipality. Each section was subsequently enriched with the respective population count. Figure \ref{fig:zones} shows the cartographic representation of these zones together with the respective population size.
\paragraph{Meteorological and Infrastructure Data:} 
\textcolor{black}{We retrieved meteorological variables and road network topologies from the ARPAV \cite{arpav2026meteo} and OpenStreetMap \cite{osm2026padua} data, respectively.} The meteorological dataset includes hourly temperature and humidity measurements for the observation period. No specific pre-processing was required for either of these two data sources, and a comprehensive description of their record structures is provided in Section \ref{sec:data_records}.

\subsection{Trajectory-Based Flow Statistics}
\label{sec:flow}
ALPR data collected by AVI devices enable a fine-grained characterization of traffic flow throughout the road network by tracking individual vehicle passages across multiple sensors. However, even when hash-encrypted, storing and releasing this information for prolonged periods can raise legitimate privacy concerns. To mitigate this risk, we developed and implemented a set of aggregated statistics derived from the ALPR recordings. This approach preserves flow information while avoiding the storage and release of individual trajectories. The mechanism involves storing hashed license plates in a buffer for a certain period of time (e.g., one day). Within this window, the hashes are used to reconstruct trajectories and compute flow-based statistics. The buffer is then flushed before the next window begins, ensuring that sensitive individual-level data are removed.
\\\\
More formally, each hash code $h$ was associated with the sequence of sensors through which it passed during the storage period. We represent this sequence of observations as a vector $[s^h_1, s^h_2, ..., s^h_k]$, together with the corresponding timestamps $[t^h_1, t^h_2, ..., t^h_k]$. We then employed the trip-division algorithm proposed in \cite{wang2023city} to partition the observations associated with each hash code $h$ into multiple trajectories. The algorithm groups consecutive detections into the same trajectory if the following condition on the temporal gap holds: $\Delta t^h = t^h_{i+1} - t^h_i < d(s_{i+1}, s_i) / v_{min}$, where $d(s_{i+1}, s_i)$ is the road distance between the two sensors, and $v_{min}$ is a minimum velocity constant set to 10 km/h (see Supplementary Section \ref{sec:ablation_vmin} for a sensitivity analysis on this parameter); otherwise, a new trajectory begins from $s_{i+1}$. To compute $d(s_{i+1}, s_i)$, we derived the shortest-path road distance between $s_i$ and $s_{i+1}$ according to the underlying road network. Each vehicle $h$ may then generate $M$ different trajectories:
\[
\mathcal{P}^h = \{\rho_m^h\}_{m=1}^M,
\]
where each trajectory $\rho^h_m$ is an ordered sequence of detections:
\[
\rho^h_m = \big( [s^h_{m,1}, s^h_{m,2}, \ldots, s^h_{m,k_m}], [t^h_{m,1}, t^h_{m,2}, \ldots, t^h_{m,k_m}] \big).
\]
Given these vehicle trajectories, we extracted the following aggregated statistics at an hourly frequency, assigning each observed transition $(s_i \rightarrow s_j)$ to the hourly time bin corresponding to the arrival time at the downstream sensor $s_j$. The hourly frequency was chosen to ensure a sufficient number of observed vehicle transitions per sensor pair within each time bin, yielding statistically stable estimates of flow statistics. Denoting with $\tau$ the hourly time bin, the statistics are defined as follows:
\begin{itemize}
    \item \textbf{Transition probability matrix:} Let $\mathcal{V}$ be the set of all sensors. For each 
    pair of sensors $(s_i, s_j) \in \mathcal{V} \times \mathcal{V}$, let $C_\tau(s_i, s_j) := \#\{\text{transitions } 
    (s_i \rightarrow s_j) \mid t_{s_j} \in \tau\}$ be the number of vehicles traveling directly from $s_i$ to 
    $s_j$ within time bin $\tau$. The outflow from sensor $s_i$ during $\tau$ is defined as the total number of vehicles departing from $s_i$ to any other sensor:
    \begin{equation}
        O_\tau(s_i) = \sum_{s_k \in \mathcal{V}} C_\tau(s_i, s_k).
        \label{eq:outflow}
    \end{equation}
    The transition probability matrix is then defined as:
    \begin{equation}
        p_{\tau}(s_i, s_j) = \frac{C_\tau(s_i, s_j)}{O_\tau(s_i)}.
        \label{prob_matrix}
    \end{equation}

    \item \textbf{Average travel time matrix:} For each pair of sensors $(s_i, s_j)$, we compute the average 
    time, in seconds, that vehicles take to travel from $s_i$ to $s_j$ during time bin $\tau$. Formally, for 
    each consecutive detection in a vehicle trajectory, define the travel time as $\Delta t_{s_i, s_j} = t_{s_j} - t_{s_i}$. 
    For each pair $(s_i, s_j)$, let the set of all observed travel times within $\tau$ be 
    $\mathcal{S}_{s_i, s_j}(\tau) = \{\Delta t_{s_i, s_j} \mid t_{s_j} \in \tau\}$. The average travel time is then defined as:
    \begin{equation}
    \mu_\tau(s_i, s_j) = \frac{1}{|\mathcal{S}_{s_i, s_j}(\tau)|} \sum_{\Delta t \in \mathcal{S}_{s_i, s_j}(\tau)} \Delta t.
    \label{eq:travel_times}
    \end{equation}

    \item \textbf{Flow residuals:} Let $I_\tau(s_i)$ be the inflow to sensor $s_i$ during time bin $\tau$, 
    defined as the total number of vehicles arriving at $s_i$ from any other sensor:
    \begin{equation}
        I_\tau(s_i) = \sum_{s_k \in \mathcal{V}} C_\tau(s_k, s_i).
    \end{equation}
    The normalized flow residual $\ell_\tau(s_i)$ is then computed for each sensor $s_i$ as:
    \begin{equation}
    \ell_\tau(s_i) = \frac{r_\tau(s_i)}{\max(I_\tau(s_i), O_\tau(s_i))}, \quad r_\tau(s_i) = I_\tau(s_i) - O_\tau(s_i).
    \label{eq:flow_residual}
    \end{equation}
\end{itemize}
We also computed a static version of these flow-based statistics by considering vehicle trajectories collected over the first 80\% of the observation period, thereby providing a summary of average traffic flow behavior. This aggregation was performed over the first 80\% of the entire period to follow standard machine learning best practices, ensuring that future data-driven models can leverage these statistics without introducing data leakage from the evaluation period.
\\\\
Note that these flow-based statistics do not provide information on complete vehicle trajectories or OD flows. Since sensors are spatially sparse, a vehicle may travel through portions of the network without being detected, and we may therefore miss substantial parts of its path. Consequently, our estimates rely on consecutive detections across sensors, i.e., observed transitions between two sensors, and provide local information on average traffic flow patterns rather than a full reconstruction of end-to-end trips.

\subsection{Linking contextual information to AVI sensors}
\label{sec:feature_enrichment}
In this section, we describe the computations performed by the Python code that we release for feature enrichment, which automatically loads traffic volume time series and performs node-level feature enrichment under a customizable configuration. By default, for each sensor $s_i$, it computes the number of POIs in each category located within a 500-meter radius, leading to a vector $\mathbf{c}_{s_i}^{\text{POI}} \in \mathbb{R}^{F_{\text{POI}}}$, where $F_{\text{POI}}$ is the number of POI categories. In addition, for each road category, it calculates its total length in meters within the same 500-meter radius around each sensor, similarly to what proposed in \cite{li2025cross}. This yields a vector $\mathbf{c}^{\text{road}}_{s_i} \in \mathbb{R}^{F_{\text{road}}}$, where $F_{\text{road}}$ is the number of road categories. For demographic information, each sensor is associated with the population of its top-$k$ ($k=5$ by default) nearest areas, determined by computing Euclidean distance between the sensor location and the centroids of its surrounding zones, resulting in a vector $\mathbf{c}^{\text{demo}}_{s_i} \in \mathbb{R}^{k}$. Meteorological variables, on the other hand, were collected from a single station and therefore represent a global, city-level feature rather than a node-specific one. The three localized feature vectors are finally concatenated into a unified context vector $\mathbf{c}_{s_i} = \left[\mathbf{c}^{\text{POI}}_{s_i} \| \mathbf{c}^{\text{road}}_{s_i} \| \mathbf{c}^{\text{demo}}_{s_i}\right] \in \mathbb{R}^{F_c}$, where $\|$ denotes the concatenation and $F_c = F_{\text{POI}} + F_{\text{road}} + k$. \textcolor{black}{Note that the released code performs this enrichment on the fly, directly from the raw contextual data. Users can therefore easily adjust the default hyper-parameters, such as the radius and the number $k$ of nearest zones, to suit their needs.}
\\\\
Building on the trajectory-derived statistics described in Section \ref{sec:flow}, we can naturally link flow information between traffic sensors using a graph-based formalism. A graph is defined as a tuple $\mathcal{G} = (\mathcal{V}, \mathcal{E})$, where $\mathcal{V}$ denotes the set of nodes and $\mathcal{E} \subseteq \mathcal{V} \times \mathcal{V}$ is the set of (directed) edges connecting pairs of nodes. In our setting, each node corresponds to a sensor, and a directed edge $(s_i, s_j) \in \mathcal{E}$ represents the presence of vehicle movements from sensor $s_i$ to $s_j$. Edge weights are derived from the aggregated trajectory information and can be defined either as transition probabilities $p_\tau(s_i,s_j)$ (Equation \eqref{prob_matrix}), capturing the likelihood that vehicles move from $s_i$ to $s_j$, or as average travel times $\mu_\tau(s_i,s_j)$ (Equation \eqref{eq:travel_times}), representing the expected travel time between the two sensors. Since these quantities are computed dynamically, both the graph topology and edge weights evolve over time. This is because a directed edge $(s_i, s_j)$ is only present in a given time bin if at least one vehicle transition from $s_i$ to $s_j$ was observed within that hour. This means that routes observed during morning rush hours may be absent in off-peak periods and vice versa, leading to different adjacency structures.
\\\\
More formally, each sensor $s_i \in \mathcal{V}$ is associated with a traffic volume measurement $x_{s_i}(t) \in \mathbb{R}$ at time $t$. Let $\mathbf{x}(t) = \left[x_{s_1}(t), \dots, x_{s_N}(t)\right]^\top \in \mathbb{R}^{N}$ denote the vector of traffic observations over all nodes at time $t$, where $t$ indexes 10-minute intervals. For the dynamic graph structure, let $\{\tau_1, \tau_2, \ldots, \tau_K\}$ denote the discrete set of hourly time bins at which the graph topology is estimated. To avoid data-leakage, each traffic observation at time $t$ is associated with the graph structure $\mathcal{G}_{\tau_k}$ estimated at the most recent preceding hour, i.e., $\tau_k = \max\{\tau \in \{\tau_1, \ldots, \tau_K\} : \tau + \Delta \leq t\}$, where $\Delta = \text{1 hour}$ is the bin duration. In this way, no look-ahead bias is introduced, as the statistics for the bin $[\tau_k, \tau_k + \Delta]$ are only available once the full hour has elapsed.

\section{Data Records} 
\label{sec:data_records}
\textcolor{black}{The dataset described in this Data Descriptor is deposited in the Zenodo
repository \cite{cappi2026padua_data}.} We release files containing time series at both the individual sensor level and in spatially aggregated form (junction-level clusters), as described in Section~\ref{sec:ts-pre-proc}, as well as the files related to contextual data sources.

\subsection{Traffic monitoring sensors} 
We release CSV files containing traffic volumes aggregated both temporally and spatially. These files have as indices the 10-minute interval timestamps, while the columns correspond to node (junction) IDs. Junction IDs follow the notation \textit{agg\_0} for the first junction, \textit{agg\_1} for the second, and so on. \textcolor{black}{A snapshot of the CSV file containing the junction-level time series data is shown in Table \ref{tab:ts_excerpt}}. All timestamps are localized to the \texttt{Europe/Rome} time zone. Additional sensors (junctions) metadata are stored in GeoPandas dataframes with the structure reported in Table~\ref{tab:sensor_metadata_fields}. 
\begin{table}[h!]
\centering
\begin{subtable}{\textwidth}
\centering
\begin{tabularx}{\textwidth}{@{}l X l@{}}
\toprule
\textbf{Field ID} & \textbf{Description} & \textbf{Data type} \\
\midrule
\textit{timestamp} & Index of the file: left edge of the 10-minute aggregation bin, localized to \texttt{Europe/Rome} (UTC+1 before the DST transition, UTC+2 after it). & datetime \\
\textit{agg\_0} $\ldots$ \textit{agg\_39} & One column per junction: number of vehicle passages recorded at that junction within the 10-minute bin. & float \\
\bottomrule
\end{tabularx}
\caption{}
\end{subtable}

\vspace{0.45cm}

\begin{subtable}{\textwidth}
\centering
\small
\begin{tabular}{lrrrrrcr}
\toprule
\textbf{timestamp} & \textbf{agg\_0} & \textbf{agg\_1} & \textbf{agg\_2} & \textbf{agg\_3} & \textbf{agg\_4} & $\cdots$ & \textbf{agg\_39} \\
\midrule
2026-02-06 01:00+01:00 & 19.0 &  6.0 & 3.0 & 8.0 & 4.0 & $\cdots$ &  1.0 \\
2026-02-06 01:10+01:00 &  7.0 &  9.0 & 0.0 & 4.0 & 3.0 & $\cdots$ & 10.0 \\
2026-02-06 01:20+01:00 &  8.0 & 19.0 & 2.0 & 6.0 & 0.0 & $\cdots$ &  4.0 \\
2026-02-06 01:30+01:00 &  9.0 & 17.0 & 3.0 & 2.0 & 0.0 & $\cdots$ &  2.0 \\
2026-02-06 01:40+01:00 &  6.0 & 13.0 & 4.0 & 5.0 & 4.0 & $\cdots$ & 10.0 \\
\bottomrule
\end{tabular}
\caption{}
\end{subtable}
\caption{\textcolor{black}{Traffic volume time series files. (a) Description of the fields; (b) snapshot of the released CSV, in which rows are indexed by 10-minute timestamps and the 40 columns \textit{agg\_0}--\textit{agg\_39} report the vehicle count at each junction. Only the first five and the last column are shown.}}
\label{tab:ts_excerpt}
\end{table}

\begin{table}[h!]
\centering
\begin{subtable}{\textwidth}
\centering
\begin{tabularx}{\textwidth}{@{}l X l@{}}
\toprule
\textbf{Field ID} & \textbf{Description} & \textbf{Data type} \\
\midrule
\textit{id} & Sensor (junction) ID. & string \\
\textit{reg\_dir} & Comma-separated list of the cardinal directions monitored at the location. & string (categorical) \\
\textit{latitude} & Sensor (junction centroid) latitude (EPSG:4326). & float \\
\textit{longitude} & Sensor (junction centroid) longitude (EPSG:4326). & float \\
\textit{geometry} & Sensor (junction centroid) location as a point (EPSG:4326). & geometry (Point) \\
\bottomrule
\end{tabularx}
\caption{}
\end{subtable}

\vspace{0.45cm}

\begin{subtable}{\textwidth}
\centering
\small
\begin{tabular}{llrrl}
\toprule
\textbf{id} & \textbf{reg\_dir} & \textbf{latitude} & \textbf{longitude} & \textbf{geometry} \\
\midrule
agg\_0 & NorthWest,SouthEast  & 45.446143 & 11.893692 & POINT (11.89369 45.44614) \\
agg\_1 & NorthSouth,SouthNorth   & 45.434895 & 11.897075 & POINT (11.89708 45.43490) \\
agg\_2 & EastWest,WestEast & 45.434708 & 11.896875 & POINT (11.89688 45.43471) \\
agg\_3 & NorthSouth,SouthNorth   & 45.430300 & 11.898690 & POINT (11.89869 45.43030) \\
agg\_4 & EastWest,WestEast & 45.430328 & 11.898653 & POINT (11.89865 45.43033) \\
\bottomrule
\end{tabular}
\caption{}
\end{subtable}
\caption{\textcolor{black}{Metadata of the sensors (junctions), released as a GeoPandas dataframe. (a) Description of the fields; (b) snapshot of the junction-level dataframe, showing the first five of the 40 junctions. The sensor-level dataframe shares the same structure over the 70 retained AVI devices.}}
\label{tab:sensor_metadata_fields}
\end{table}
\paragraph{Direction field semantics:} For AVI sensors, valid \textit{direction} values are \textit{NorthSouth, SouthNorth, NorthWest, NorthEast, SouthWest, SouthEast, EastWest, WestEast}. Each label identifies the direction monitored by the sensor, i.e., it should be read as ``towards South" (\textit{NorthSouth}), ``towards North" (\textit{SouthNorth}), ``towards East" (\textit{WestEast}), etc., rather than as an origin--destination pair (``from North to South"). Diagonal values (e.g., \textit{NorthEast}) are interpreted analogously as ``towards North-East".

\subsubsection{Aggregated flow statistics} 
The CSV files containing dynamic transition probability matrices, average travel times and flow residuals follow the structure reported in Tables \ref{tab:flow_transition_fields}, \ref{tab:flow_traveltime_fields}, and \ref{tab:flow_residual_fields}, respectively. In all tables, the \textit{time\_bin} field is expressed in local time, i.e., UTC+1 before the DST transition and UTC+2 after it. To compute the flow statistics with respect to the junction-aggregated version of the dataset, we merged the trajectories of vehicles passing through sensors at the same intersection into a single passage through the junction. The files reporting flow-based statistics aggregated over the first 80\% of the observation period share the same structure as the aforementioned dynamic files, except for the \textit{time\_bin} field.
\begin{table}[h!]
\centering
\begin{subtable}{\textwidth}
\centering
\begin{tabularx}{\textwidth}{@{}l X l@{}}
\toprule
\textbf{Field ID} & \textbf{Description} & \textbf{Data type} \\
\midrule
\textit{time\_bin} & Timestamp with hourly frequency (UTC+1, UTC+2). & datetime \\
\textit{from} & ID of the source sensor (junction). & string \\
\textit{to} & ID of the target sensor (junction). & string \\
\textit{count\_ij} & Number of vehicles traveling directly along the \textit{from-to} segment within \textit{time\_bin}. & float \\
\textit{count\_i} & Outflow $O_\tau(s_i)$ (Eq. \eqref{eq:outflow}) from the source sensor (junction) within \textit{time\_bin}. & float \\
\textit{P\_ij} & Probability of observing a \textit{from-to} segment within \textit{time\_bin}, i.e., \textit{count\_ij} / \textit{count\_i}. & float \\
\bottomrule
\end{tabularx}
\caption{}
\end{subtable}

\vspace{0.45cm}

\begin{subtable}{\textwidth}
\centering
\small
\begin{tabular}{lllrrr}
\toprule
\textbf{time\_bin} & \textbf{from} & \textbf{to} & \textbf{count\_ij} & \textbf{count\_i} & \textbf{P\_ij} \\
\midrule
2026-02-06 01:00:00+01:00 & agg\_1 & agg\_0  & 13.0 & 17.0 & 0.764706 \\
2026-02-06 01:00:00+01:00 & agg\_1 & agg\_11 &  1.0 & 17.0 & 0.058824 \\
2026-02-06 01:00:00+01:00 & agg\_1 & agg\_3  &  1.0 & 17.0 & 0.058824 \\
2026-02-06 01:00:00+01:00 & agg\_1 & agg\_31 &  1.0 & 17.0 & 0.058824 \\
2026-02-06 01:00:00+01:00 & agg\_1 & agg\_6  &  1.0 & 17.0 & 0.058824 \\
\bottomrule
\end{tabular}
\caption{}
\end{subtable}
\caption{\textcolor{black}{Dynamic transition probability matrix files. (a) Description of the fields; (b) snapshot of the junction-level file, reporting the transitions observed from junction \textit{agg\_1} in the first hourly bin of the observation period.}}
\label{tab:flow_transition_fields}
\end{table}

\begin{table}[h!]
\centering
\begin{subtable}{\textwidth}
\centering
\begin{tabularx}{\textwidth}{@{}l X l@{}}
\toprule
\textbf{Field ID} & \textbf{Description} & \textbf{Data type} \\
\midrule
\textit{time\_bin} & Timestamp with hourly frequency (UTC+1, UTC+2). & datetime \\
\textit{from} & ID of the source sensor (junction). & string \\
\textit{to} & ID of the target sensor (junction). & string \\
\textit{median} & Median travel time between \textit{from} and \textit{to} within \textit{time\_bin}. & float (seconds) \\
\textit{mean} & Average travel time $\mu_\tau(s_i,s_j)$ (Eq. \eqref{eq:travel_times}) between \textit{from} and \textit{to} within \textit{time\_bin}. & float (seconds) \\
\textit{sum} & Sum of the observed travel times between \textit{from} and \textit{to} within \textit{time\_bin}. & float (seconds) \\
\textit{count} & Number of observed transitions used to compute the statistics. & float \\
\bottomrule
\end{tabularx}
\caption{}
\end{subtable}

\vspace{0.45cm}

\begin{subtable}{\textwidth}
\centering
\small
\setlength{\tabcolsep}{4pt}
\begin{tabular}{lllrrrr}
\toprule
\textbf{time\_bin} & \textbf{from} & \textbf{to} & \textbf{median} & \textbf{mean} & \textbf{sum} & \textbf{count} \\
\midrule
2026-02-06 01:00:00+01:00 & agg\_1 & agg\_0  & 104.607 & 108.937 & 1416.185 & 13.0 \\
2026-02-06 01:00:00+01:00 & agg\_1 & agg\_11 &  60.783 &  60.783 &   60.783 &  1.0 \\
2026-02-06 01:00:00+01:00 & agg\_1 & agg\_3  & 166.411 & 166.411 &  166.411 &  1.0 \\
2026-02-06 01:00:00+01:00 & agg\_1 & agg\_31 & 640.088 & 640.088 &  640.088 &  1.0 \\
2026-02-06 01:00:00+01:00 & agg\_1 & agg\_6  & 107.516 & 107.516 &  107.516 &  1.0 \\
\bottomrule
\end{tabular}
\caption{}
\end{subtable}
\caption{\textcolor{black}{Dynamic average travel times matrix files. (a) Description of the fields; (b) snapshot of the junction-level file, reporting the travel times observed from junction \textit{agg\_1} in the first hourly bin of the observation period.}}
\label{tab:flow_traveltime_fields}
\end{table}

\begin{table}[h!]
\centering
\begin{subtable}{\textwidth}
\centering
\begin{tabularx}{\textwidth}{@{}l X l@{}}
\toprule
\textbf{Field ID} & \textbf{Description} & \textbf{Data type} \\
\midrule
\textit{time\_bin} & Timestamp with hourly frequency (UTC+1, UTC+2). & datetime \\
\textit{junction} (\textit{node\_id}) & Junction (sensor) ID. & string \\
\textit{inflow} & Sensor's input flow $I_\tau(s_i)$. & float \\
\textit{outflow} & Sensor's output flow $O_\tau(s_i)$ (Eq. \eqref{eq:outflow}). & float \\
\textit{residual} & Flow residual \textit{inflow} $-$ \textit{outflow}. & float \\
\textit{normalized\_residual} & Normalized residual computed as described in Eq. \eqref{eq:flow_residual}. & float \\
\bottomrule
\end{tabularx}
\caption{}
\end{subtable}

\vspace{0.45cm}

\begin{subtable}{\textwidth}
\centering
\small
\setlength{\tabcolsep}{4pt}
\begin{tabular}{llrrrr}
\toprule
\textbf{time\_bin} & \textbf{junction} & \textbf{inflow} & \textbf{outflow} & \textbf{residual} & \textbf{normalized\_residual} \\
\midrule
2026-02-06 01:00:00+01:00 & agg\_0  & 22.0 &  0.0 &  22.0 &  1.000000 \\
2026-02-06 01:00:00+01:00 & agg\_1  & 38.0 & 17.0 &  21.0 &  0.552632 \\
2026-02-06 01:00:00+01:00 & agg\_10 & 15.0 &  3.0 &  12.0 &  0.800000 \\
2026-02-06 01:00:00+01:00 & agg\_11 &  5.0 & 12.0 &  -7.0 & -0.583333 \\
2026-02-06 01:00:00+01:00 & agg\_12 &  5.0 &  3.0 &   2.0 &  0.400000 \\
\bottomrule
\end{tabular}
\caption{}
\end{subtable}
\caption{\textcolor{black}{Dynamic flow residuals files. (a) Description of the fields; the identifier column is named \textit{junction} in the junction-level file and \textit{node\_id} in the sensor-level one; (b) snapshot of the junction-level file, reporting the first five junctions in the first hourly bin of the observation period.}}
\label{tab:flow_residual_fields}
\end{table}

\subsection{Contextual information}
POIs and meteorological data are organized in CSV files with the structures detailed in Tables \ref{tab:poi_fields} and \ref{tab:meteo_fields}, respectively, while demographic data are stored in a GeoPandas DataFrame with the structure detailed in Table \ref{tab:demo_fields}. Note that, when performing the feature enrichment described in Section \ref{sec:feature_enrichment}, all geometries are projected to EPSG:3003,  ensuring that the radius around each sensor, expressed in meters, can be correctly computed. Finally, the road network is stored as a directed graph in \texttt{graphml} format.
\begin{table}[h!]
\centering
\begin{subtable}{\textwidth}
\centering
\begin{tabularx}{\textwidth}{@{}l X l@{}}
\toprule
\textbf{Field ID} & \textbf{Description} & \textbf{Data type} \\
\midrule
\textit{id} & Unique ID for each POI. & string \\
\textit{confidence} & Confidence parameter for the existence of the POI. & float \\
\textit{primary\_category} & Fine-grained category. & string \\
\textit{longitude} & POI longitude (EPSG:4326). & float \\
\textit{latitude} & POI latitude (EPSG:4326). & float \\
\textit{category\_level\_0} & POI category after hierarchical reclassification (see Section \ref{sec:context}). & string \\
\bottomrule
\end{tabularx}
\caption{}
\end{subtable}

\vspace{0.45cm}

\begin{subtable}{\textwidth}
\centering
\footnotesize
\setlength{\tabcolsep}{4pt}
\begin{tabular}{lrlrrl}
\toprule
\textbf{id} & \textbf{confidence} & \textbf{primary\_category} & \textbf{longitude} & \textbf{latitude} & \textbf{category\_level\_0} \\
\midrule
d7b9ddae$\ldots$c1e5 & 0.770 & commercial\_real\_estate    & 11.8770 & 45.4119 & real\_estate \\
bbc247a2$\ldots$0e85 & 0.770 & fashion\_accessories\_store & 11.8566 & 45.3831 & retail \\
d1a7c9d8$\ldots$0a0e & 0.950 & asian\_restaurant          & 11.8781 & 45.4169 & eat\_and\_drink \\
3cf10b4e$\ldots$3862 & 0.950 & college\_university        & 11.8798 & 45.3964 & education \\
6130fd84$\ldots$eda0 & 0.995 & professional\_services     & 11.9010 & 45.4097 & professional\_services \\
\bottomrule
\end{tabular}
\caption{}
\end{subtable}
\caption{\textcolor{black}{POIs files. (a) Description of the fields; (b) snapshot of the released CSV, showing five of the 9,262 retained POIs spanning the range of confidence values. The \textit{id} values are UUIDs and are abbreviated here for readability.}}
\label{tab:poi_fields}
\end{table}

\begin{table}[h!]
\centering
\begin{subtable}{\textwidth}
\centering
\begin{tabularx}{\textwidth}{@{}l X l@{}}
\toprule
\textbf{Field ID} & \textbf{Description} & \textbf{Data type} \\
\midrule
\textit{id} & ID of the census section. & int \\
\textit{population} & Number of inhabitants of the corresponding section. & int \\
\textit{shape\_area} & Area covered by the shape of the section. & float (m$^2$) \\
\textit{geometry} & Census section shape as a polygon (EPSG:3003). & geometry (Polygon) \\
\bottomrule
\end{tabularx}
\caption{}
\end{subtable}

\vspace{0.45cm}

\begin{subtable}{\textwidth}
\centering
\small
\begin{tabular}{rrrl}
\toprule
\textbf{id} & \textbf{population} & \textbf{shape\_area} & \textbf{geometry} \\
\midrule
2 & 110 & 11132.818669 & POLYGON ((1724992.875 5032282.788, $\ldots$)) \\
3 & 167 & 12278.580369 & POLYGON ((1724991.917 5032290.121, $\ldots$)) \\
4 &  98 & 10897.610366 & POLYGON ((1725017.375 5032290.538, $\ldots$)) \\
5 &  47 &  6561.765099 & POLYGON ((1725209.536 5032354.408, $\ldots$)) \\
6 &  37 &  2753.646036 & POLYGON ((1725169.065 5032353.538, $\ldots$)) \\
\bottomrule
\end{tabular}
\caption{}
\end{subtable}
\caption{\textcolor{black}{Demographic data, released as a GeoPandas dataframe. (a) Description of the fields; (b) snapshot showing five of the 1,777 census sections. Polygon coordinates are truncated after the first vertex for readability.}}
\label{tab:demo_fields}
\end{table}

\begin{table}[h!]
\centering
\begin{subtable}{\textwidth}
\centering
\begin{tabularx}{\textwidth}{@{}l X l@{}}
\toprule
\textbf{Field ID} & \textbf{Description} & \textbf{Data type} \\
\midrule
\textit{timestamp} & Timestamp with hourly frequency (UTC+1, UTC+2). & datetime \\
\textit{avg\_temp} & Average temperature. & float ($^\circ$C) \\
\textit{prec} & Precipitation. & float (mm) \\
\textit{hum\_min} & Minimum humidity. & int (\%) \\
\textit{hum\_max} & Maximum humidity. & int (\%) \\
\bottomrule
\end{tabularx}
\caption{}
\end{subtable}

\vspace{0.45cm}

\begin{subtable}{\textwidth}
\centering
\small
\begin{tabular}{lrrrr}
\toprule
\textbf{timestamp} & \textbf{avg\_temp} & \textbf{prec} & \textbf{hum\_min} & \textbf{hum\_max} \\
\midrule
2026-02-06 01:00:00+01:00 & 7.8 & 0.0 & 100 & 100 \\
2026-02-06 02:00:00+01:00 & 7.5 & 0.0 & 100 & 100 \\
2026-02-06 03:00:00+01:00 & 7.6 & 0.0 & 100 & 100 \\
2026-02-06 04:00:00+01:00 & 7.6 & 0.2 & 100 & 100 \\
2026-02-06 05:00:00+01:00 & 7.5 & 0.4 & 100 & 100 \\
\bottomrule
\end{tabular}
\caption{}
\end{subtable}
\caption{\textcolor{black}{Meteorological variables files. (a) Description of the fields; (b) snapshot of the released CSV, showing the first five hourly records of the observation period.}}
\label{tab:meteo_fields}
\end{table}

\section{Technical Validation}
To demonstrate the quality and reliability of the presented dataset, we performed a series of analysis to verify that the collected time series reproduce well-known traffic regularities, such as the daily rhythms of rush hours and the contrast between weekday and weekend traffic regimes. Furthermore, we considered the mid-February school closure as a case study to assess whether the traffic volume dataset was sensitive enough to reflect the impact of school-related trips.
\begin{figure}[h]
    \centering
    
    \begin{subfigure}{0.55\linewidth}
        \centering
        \includegraphics[width=\linewidth]{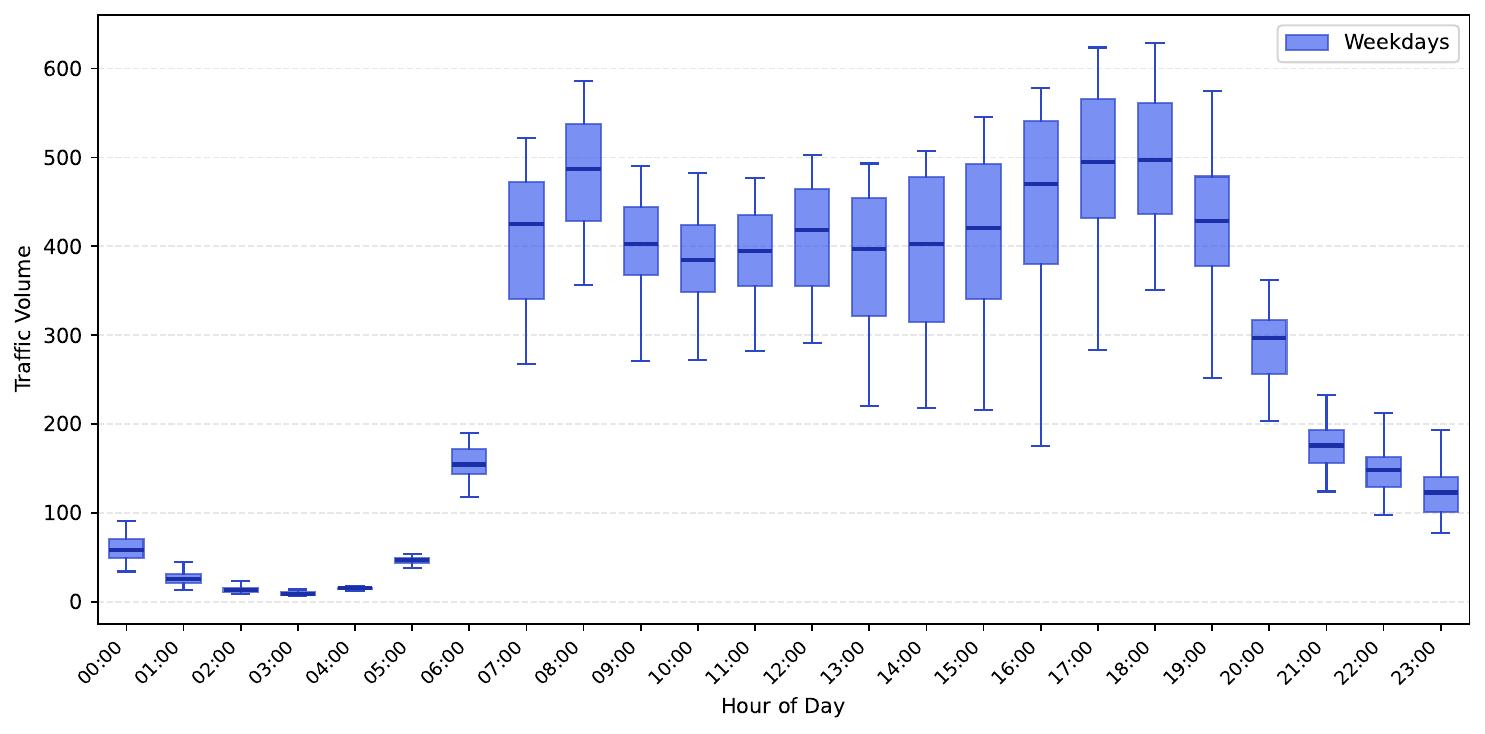}
        \caption{}
        \label{fig:weekdays}
    \end{subfigure}
    
    \begin{subfigure}{0.55\linewidth}
        \centering
        \includegraphics[width=\linewidth]{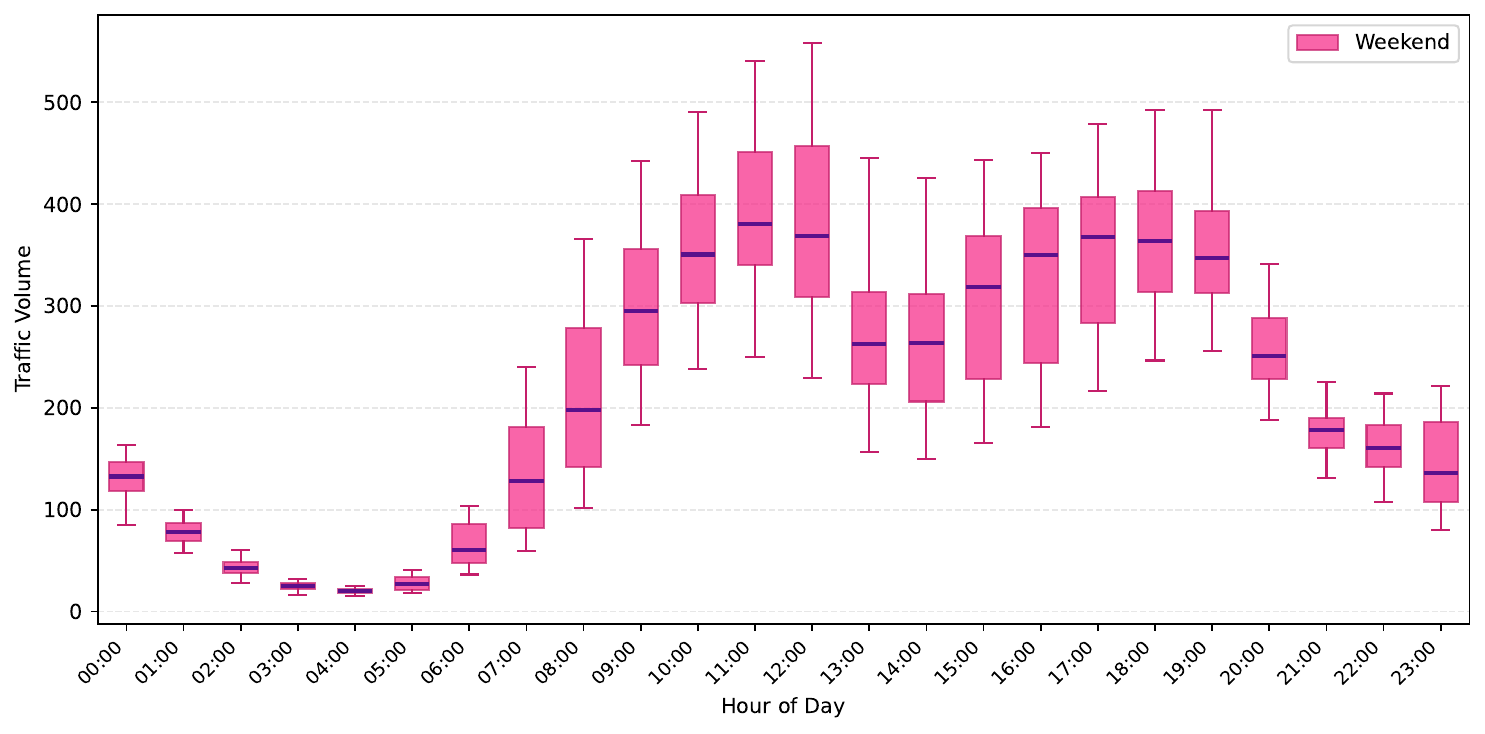}
        \caption{}
        \label{fig:weekends}
    \end{subfigure}
    
    \caption{Hourly traffic volume distribution on (a) weekdays, and (b) weekend days.}
    \label{fig:weekdays_weekends}
\end{figure}

\subsection{Traffic volume peaks}
To validate that the dataset captures the expected temporal demand patterns, we compared the hourly traffic volume distributions on weekdays and on weekend days, depicted in Figures \ref{fig:weekdays}, \ref{fig:weekends}, respectively. As expected, weekdays exhibit clear peaks in the morning around 8 AM and in the late afternoon around 5-6 PM, reflecting the incoming and return flows of commuters at the beginning and at the end of working days, respectively. In contrast, weekend days show substantially reduced morning traffic and a general flatter traffic volume profile. Specifically, the morning peak disappears, indicating the absence of work-related trips that produce weekday congestion. Instead, weekend traffic rises more gradually over the morning, reaching a plateau during midday and early afternoon hours.

\subsection{School closure}
\begin{figure}[h!]
    \centering
    \begin{subfigure}[b]{0.5\linewidth}
        \centering
        \includegraphics[width=\linewidth]{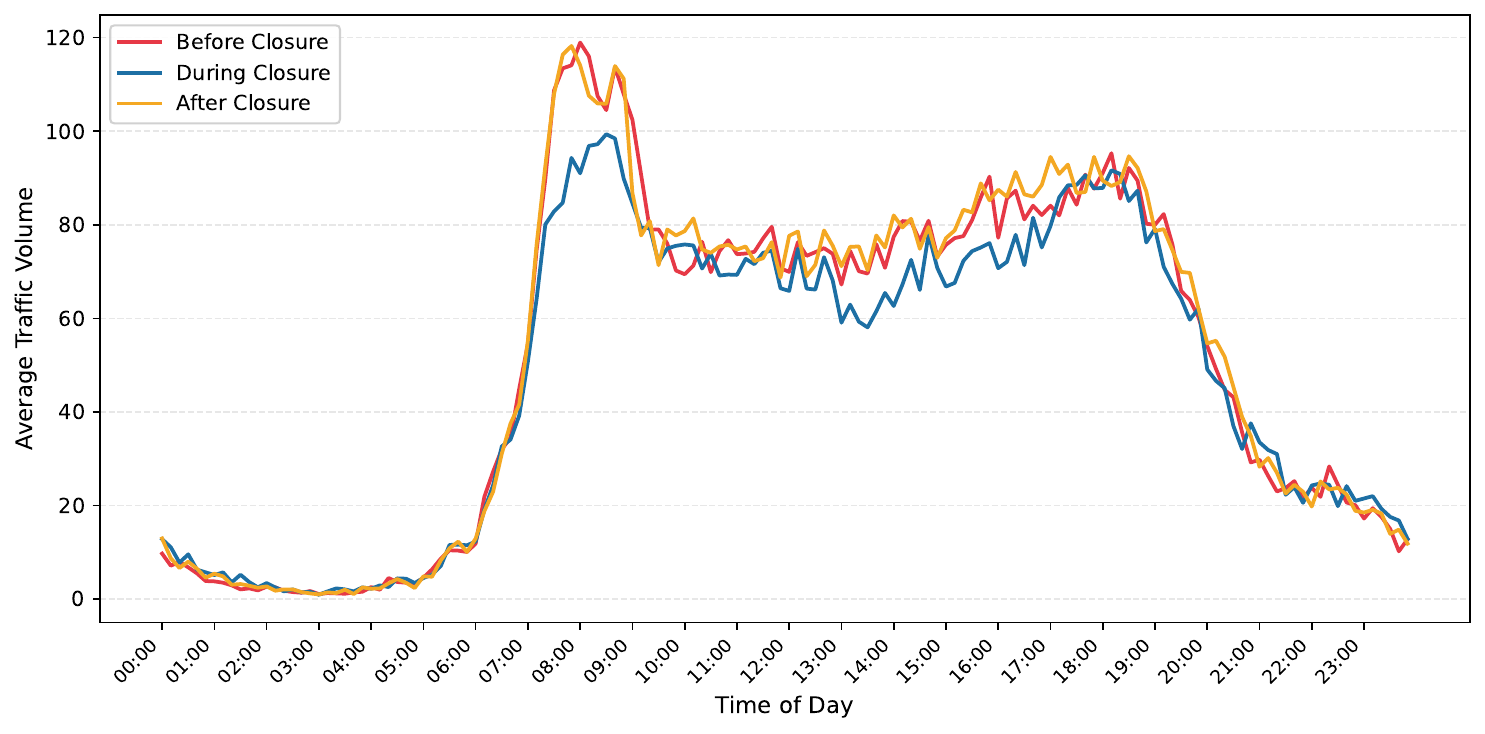}
        \caption{}
        \label{fig:school_closure_north}
    \end{subfigure}
    \hfill
    \begin{subfigure}[b]{0.4\linewidth}
        \centering
        \includegraphics[width=\linewidth]{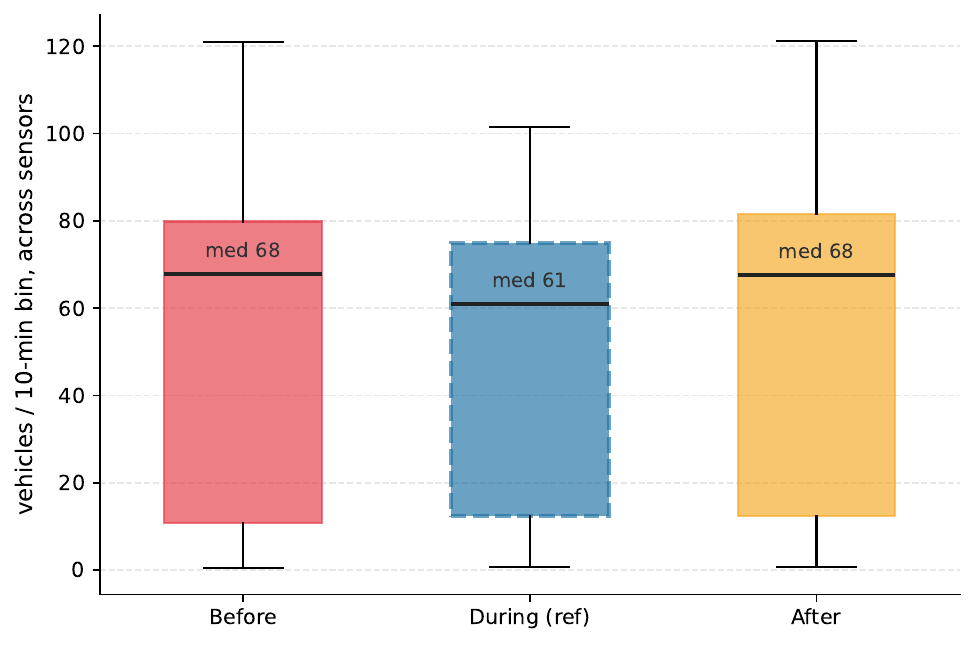}
        \caption{}
        \label{fig:school_closure_volume}
    \end{subfigure}
    \caption{ \textcolor{black}{(a) Average traffic volume profiles before, during, and after the school closure period. (b) Traffic volume distributions across sensors before, during, and after the school closure period.}
    }
    \label{fig:school_closure}
\end{figure}

To further validate the reliability of the presented dataset, we examined the average traffic volume profiles before, during, and after the school closure period that occurred from 16 to 18 February. To do so, we selected a subset of sensors located in the North-West part of the city close to several schools. The \textit{before closure} and \textit{after closure} periods correspond to Monday-through-Wednesday weekdays of the week immediately preceding and following the 16–18 February closure period, respectively. The average traffic volume profiles over the three periods is shown in Figure \ref{fig:school_closure_north}. We can observe two major drops in traffic volumes during the closure period: one in the morning around 8 AM and another in the early afternoon around 1 PM. These drops coincide with typical school schedules, with the morning one indicating the absence of traffic at opening time, while the other reflects the lack of pick-up activity at the end of the morning school session. \textcolor{black}{In Figure \ref{fig:school_closure_volume} we show the distributions of traffic volumes across sensors in the three periods. During schools closure we observe a decrease of approximately 10\% in traffic volume compared to the other two periods.}

\subsection{Zero value frequency} 
\begin{figure}[h!]
    \centering
    \includegraphics[width=0.55\linewidth]{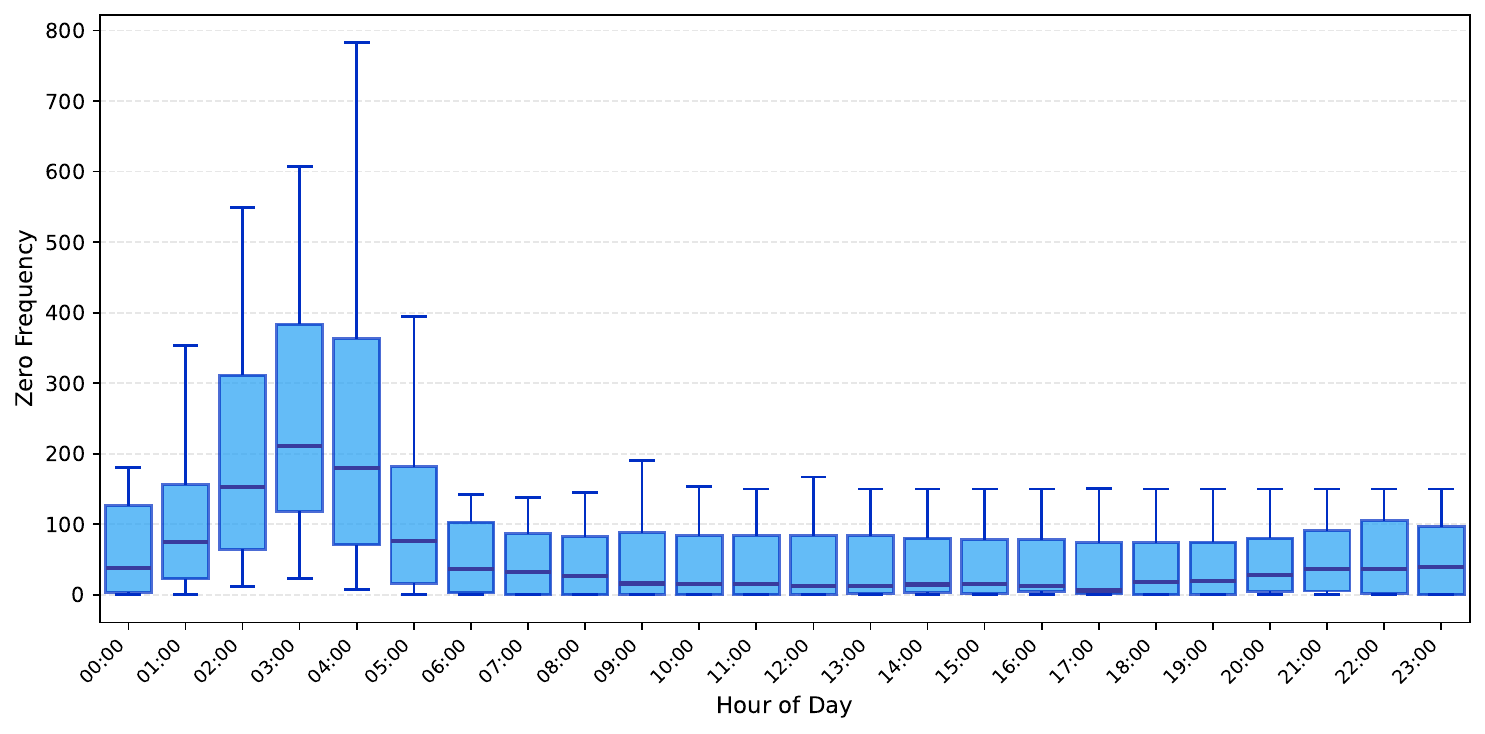}
    \caption{Zero volume frequency across hours.}
    \label{fig:zero_value_freq}
\end{figure}

Figure \ref{fig:zero_value_freq} shows the zero volume frequency distribution over hours of the day aggregated throughout the observed period. Zero counts are concentrated during the night and early morning (roughly 1 AM - 5 AM), with a peak at 3 AM, and become scarce during the daytime hours when traffic is expected to be present. The slightly positive frequency of zeros observed in the late afternoon and evening indicates the presence of a subset of low-volume sensors that report zero counts even outside the night period. In general, the analysis on traffic volume distributions and zero volume frequency demonstrates the reliability of the presented time series dataset and its ability to capture traffic fluctuations throughout the day. 
\begin{figure}[h!]
    \centering
    
    \begin{subfigure}[b]{0.48\linewidth}
    \centering
    \includegraphics[width=\linewidth]{images/travel_time_profile_weekday.png}
    \caption{}
    \label{fig:weekdays_travel}
    \end{subfigure}
    \hfill
    \begin{subfigure}[b]{0.48\linewidth}
        \centering
        \includegraphics[width=\linewidth]{images/travel_time_profile_weekend.png}
        \caption{}
        \label{fig:weekends_travel}
    \end{subfigure}

    \vspace{0.3cm}

    \begin{subfigure}[b]{0.5\linewidth}
        \centering
        \includegraphics[width=\linewidth]{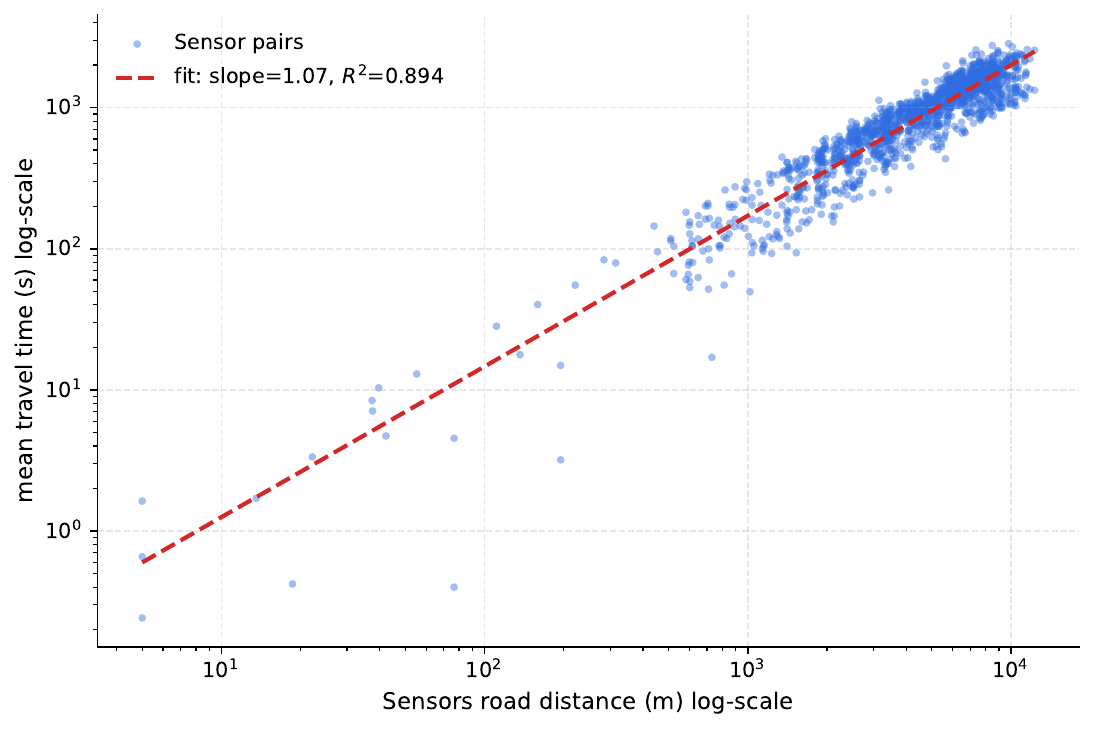}
        \caption{}
        \label{fig:travel_time_vs_distance}
    \end{subfigure}

    \caption{Hourly average travel time distributions on (a) weekdays and (b) weekend days. \textcolor{black}{Sub-figure (c) shows the aggregated average travel time against the shortest-path road distance separating each pair of sensors, on logarithmic axes, together with the least-squares fit.}
    }
    \label{fig:weekdays_weekends_travel_time}
\end{figure}

\subsection{Traffic flow statistics} 
Regarding the trajectory-based data, we show in Figures \ref{fig:weekdays_travel} and \ref{fig:weekends_travel} the average hourly travel time distribution computed throughout the observed period, comparing weekdays and weekend days. The shape of the distribution further confirms the presence of the 8 AM morning traffic peak, the early afternoon drop, and the subsequent evening peak, coinciding with the usual rush hour periods observed during weekdays. Furthermore, travel times during weekends exhibit a temporal profile very similar to the traffic volume distribution depicted in Figure \ref{fig:weekends}, without the sharp rush-hour peaks observed on weekdays. Notably, average travel times during night are higher on weekend days than on weekdays, providing further insights into nighttime activities specific to weekend behaviors. \textcolor{black}{As a further consistency check, Figure \ref{fig:travel_time_vs_distance} compares the statically aggregated average travel time of each observed sensor pair with the shortest-path road distance separating the two sensors. Both quantities are shown on logarithmic axes. The least-squares fit provides an $R^2 = 0.894$, confirming, as expected, that the two quantities are linearly correlated. We refer to Supplementary Section \ref{app:flow_analyses} for additional analyses on flow-based statistics.}
\section{Usage Notes}
The dataset is released together with a Python class that facilitates loading and pre-processing of traffic volume time series and their associated contextual features. The following snippet illustrates how to instantiate the \texttt{TrafficData} class and iterate over the resulting sliding-window dataset using a PyG \texttt{DataLoader}:
\begin{lstlisting}[language=Python]
from torch_geometric.loader import DataLoader
from data_loaders.TrafficData import TrafficData
 
dataset = TrafficData(
    root='./data',
    name='traffic_padua',
    device='cpu',
    history=6,
    horizon=6,
    stride=1,
    flow_adj=True,
    dyn_adj=False,
    flow_threshold=0.0,
    use_avg_travel_times=False,
    zero_run_threshold=6
)
 
loader = DataLoader(dataset, batch_size=32, shuffle=True)

for batch in loader:
    print(batch)
\end{lstlisting}
\noindent The constructor arguments are described below:
\begin{itemize}
    \item \texttt{root}, \texttt{name}: root directory for data storage and dataset identifier used for processed-file naming.
    \item \texttt{device}: PyTorch device string (e.g., \texttt{'cpu'}, \texttt{'cuda:0'}).
    \item \texttt{history}, \texttt{horizon}: input sequence length and forecast horizon, expressed as number of 10-minute intervals.
    \item \texttt{stride}: step size between consecutive sliding windows.
    \item \texttt{flow\_adj}: if \texttt{True}, the graph is built from vehicle-flow transition probabilities; otherwise, road-network proximity is used as in \cite{DBLP:conf/iclr/LiYS018}.
    \item \texttt{dyn\_adj}: if \texttt{True}, hourly time-varying adjacency matrices are used to construct the dynamic graph. Requires \texttt{flow\_adj=True}.
    \item \texttt{flow\_threshold}: minimum transition probability for an edge to be included in the graph.
    \item \texttt{use\_avg\_travel\_times}: if \texttt{True}, edge weights are set to average travel times instead of transition probabilities.
    \item \texttt{zero\_run\_threshold}: number of consecutive zero readings above which observations are masked as inactive sensor data. This creates two binary masks, one for the input window and one for the horizon window.
\end{itemize}

\section{Data Availability}
\textcolor{black}{The files related to the dataset described in this paper can be found in the following Zenodo repository: \url{https://doi.org/10.5281/zenodo.22111450}.}

\section{Code Availability}
\label{sec:code}
The code for the analysis is written in Python 3.12.0, and can be found at the following GitHub repository: https://github.com/riccardocappi/Traffic-Data-Padua in the \texttt{technical\_valid.ipynb} notebook. The repository contains a \texttt{README} file that further explains how to install the required packages to reproduce the analysis and how to use the Python code to automatically load time series and contextual data.

\bibliographystyle{plain} 
\bibliography{bibliography} 

\section{Author Contributions}
R. Cappi conducted the analyses, developing the code for data collection, pre-processing, and dataset organization. M. Luca and B. Lepri provided guidance on the mobility-related aspects of the work, including identifying relevant external data sources and advising on how to structure and integrate them within the dataset. R. Cappi and M. Luca wrote the manuscript. A. Sperduti and B. Lepri designed the study. P. Fontolan contributed in the data acquisition process. N. Navarin, A. Sperduti, and B. Lepri supervised the work, revised the analyses, and contributed with useful discussions.

\section{Competing Interests}
The authors declare no competing interests.

\section{Acknowledgements}
This research was done with the collaboration of the Municipality of Padua and supported by its Innovation and Digital Transition Sector. The authors wish to thank Giacomo Toto and Alberto Corò for their technical and administrative support. Some POIs and street network data are copyrighted by OpenStreetMap contributors and available from https://www.openstreetmap.org. Open Database License: © OpenStreetMap contributors, Overture Maps Foundation

\section{Funding}
R. Cappi scholarship is primarily funded by the Municipality of Padua (grant no.: N/A) and the Department of Mathematics at the University of Padua (grant no.: N/A). No grant numbers are associated with this funding, as it was provided through direct institutional support.

\clearpage                      

\setcounter{section}{0}
\setcounter{figure}{0}
\setcounter{table}{0}
\setcounter{equation}{0}

\renewcommand{\thesection}{S\arabic{section}}
\renewcommand{\thesubsection}{S\arabic{section}.\arabic{subsection}}
\renewcommand{\thefigure}{S\arabic{figure}}
\renewcommand{\thetable}{S\arabic{table}}
\renewcommand{\theequation}{S\arabic{equation}}

\section*{Supplementary Information}
\label{app:appendix}


\section{Sensitivity Analysis on $v_{\min}$}
\label{sec:ablation_vmin}

To assess the sensitivity of the trajectory-based flow statistics to the choice of the minimum velocity parameter $v_{\min}$, we computed time-varying transition probability matrices under three configurations ($v_{\min} \in \{10, 15, 20\}$ km/h) and compared them in terms of graph topology, edge-level probability estimates, and node out-degree distributions. The matrices were computed with respect to the junction-level version of the dataset.

\paragraph{Graph topology}
Table \ref{tab:jaccard} reports the pairwise edge overlap between configurations in terms of Jaccard similarity and number of shared edges. The Jaccard similarity is consistently high across all pairs, ranging from \textcolor{black}{0.908 to 0.960}, indicating that the three graphs share the vast majority of their edges.

\begin{table}[h]
\centering
\begin{tabular}{lcc}
\toprule
Comparison & Jaccard & Shared \\
\midrule
$v_{\min}{=}10$ vs.\ $v_{\min}{=}15$ & 0.960 & 4195  \\
$v_{\min}{=}10$ vs.\ $v_{\min}{=}20$ & 0.908 & 3969  \\
$v_{\min}{=}15$ vs.\ $v_{\min}{=}20$ & 0.946 & 3968 \\
\bottomrule
\end{tabular}
\caption{Pairwise edge overlap between $v_{\min}$ configurations.}
\label{tab:jaccard}
\end{table}

\paragraph{Transition probability estimates.}
Table \ref{tab:mae} reports the Mean Absolute Error (MAE) of $p_{ij}$ computed on the edges shared between each pair of configurations. The MAE remains small in all cases, with adjacent configurations ($v_{\min}{=}10$ vs.\ $v_{\min}{=}15$ and $v_{\min}{=}15$ vs.\ $v_{\min}{=}20$) differing by less than 0.02, and the most distant pair ($v_{\min}{=}10$ vs.\ $v_{\min}{=}20$) reaching a MAE of only 0.04. This indicates that the transition probability estimates are largely stable across configurations.

\begin{table}[h]
\centering
\begin{tabular}{lc}
\toprule
Comparison & MAE \\
\midrule
$v_{\min}{=}10$ vs.\ $v_{\min}{=}15$ & 0.0210 \\
$v_{\min}{=}15$ vs.\ $v_{\min}{=}20$ & 0.0232 \\
$v_{\min}{=}10$ vs.\ $v_{\min}{=}20$ & 0.0408 \\
\bottomrule
\end{tabular}
\caption{Mean Absolute Error (MAE) of $p_{ij}$ on shared edges between $v_{\min}$ configurations.}
\label{tab:mae}
\end{table}

\begin{figure}[h]
    \centering
    \includegraphics[width=0.55\linewidth]{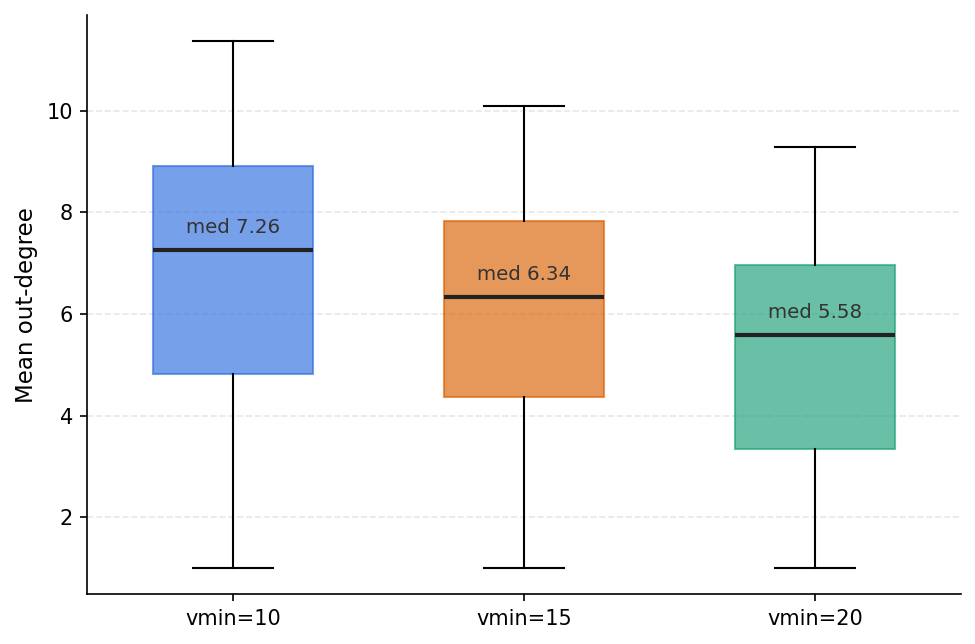}
    \caption{Mean out-degree distribution per source node for $v_{\min} \in \{10, 15, 20\}$ km/h.}
    \label{fig:ablation_outdegree}
\end{figure}

\paragraph{Out-degree distribution.}
Figure \ref{fig:ablation_outdegree} shows the mean out-degree distribution per source node across the three $v_{\min}$ values. While median out-degree decreases slightly with stricter thresholds, the quartile ranges overlap substantially, confirming that the overall graph connectivity structure is preserved.

\section{\textcolor{black}{Analyses on Traffic Volumes Invalid Observations}}
\label{app:missingness_stats}
\begin{figure}[h]
    \centering
    \includegraphics[width=0.6\linewidth]{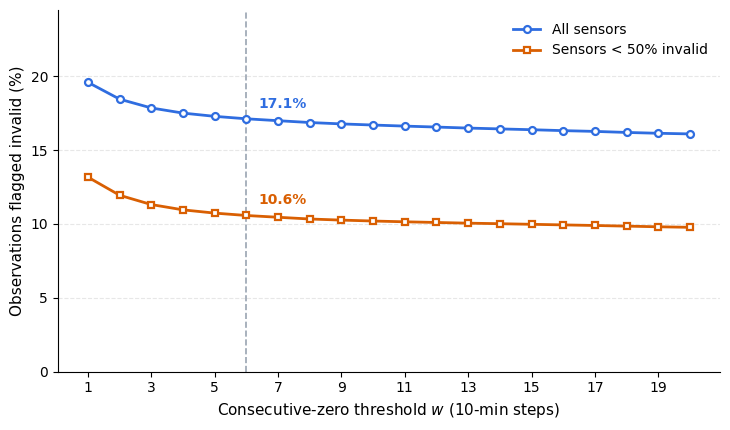}
    \caption{
    \textcolor{black}{Percentage of traffic-volume observations flagged as invalid as a function of the
    consecutive-zero threshold $w$, expressed in number of 10-minute steps. The blue line refers
    to the full set of 70 AVI sensors, the orange line to the subset obtained after discarding
    devices with more than 50\% of invalid observations at $w = 6$. The dashed vertical line marks
    $w = 6$, i.e., the one-hour threshold adopted in Section~\ref{sec:ts-pre-proc}.}}
    \label{fig:time_window_ablation}
\end{figure}

\begin{figure}[h]
    \centering
    \includegraphics[width=0.6\linewidth]{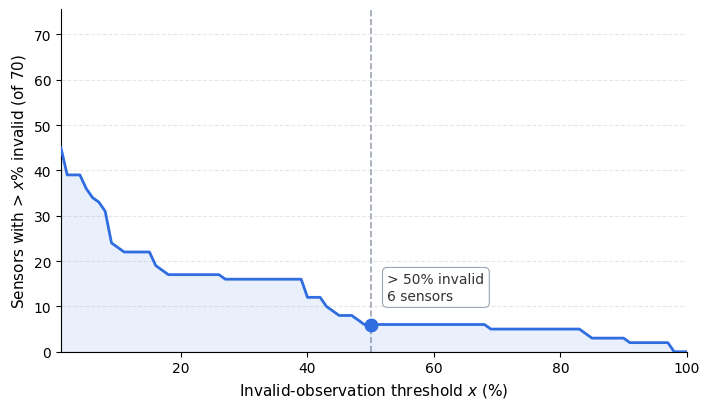}
    \caption{
    \textcolor{black}{Number of AVI sensors, out of the 70 released devices, whose fraction of invalid observations exceeds a given threshold $x$, computed with a consecutive-zero threshold of one
    hour. The dashed vertical line marks the 50\% criterion adopted in Section~\ref{sec:ts-pre-proc}.}
    }
    \label{fig:sensors_gt_threshold}
\end{figure}

\textcolor{black}{In Figure \ref{fig:time_window_ablation} we show the percentage of observations flagged as invalid by increasing the consecutive-zero threshold $w$ (see Section \ref{sec:ts-pre-proc}) from 1 to 20. This process is applied with respect to two sets of sensors: (i) the full set of 70 AVI devices (blue line), and (ii) the set of retained sensors after filtering devices with more than 50\% of observations flagged as invalid considering a consecutive-zero threshold of 1 hour (orange line). The Figure shows that in both cases the choice of the zero-run value does not substantially impact the percentage of invalid observations. Over the range of $w$, the fraction of flagged measurements varies from 19.59\% to 16.10\% for the full set of devices, and from 13.17\% to 9.77\% for the filtered one, i.e., less than four percentage points in both cases. Furthermore, both curves flatten for $w \geq 6$ (1 hour) indicating that most of the masked observations belong to long zero runs attributable to sensor inactivity. The gap between the two curves, roughly six percentage points at every threshold, shows that the invalid observations are not uniformly distributed across devices but concentrated in a small group of malfunctioning sensors.
} 
\textcolor{black}{Figure \ref{fig:sensors_gt_threshold} reports, for a fixed 1-hour $w$ threshold, the number of sensors whose fraction of invalid observations exceeds a given value $x$, with $x$ varying from 0\% to 100\%. The curve
decreases sharply for small values of $x$ and stabilizes up to $x \approx 40\%$ before dropping to 6 sensors at the $50\%$ criterion adopted in Section~\ref{sec:ts-pre-proc}. The remaining tail decays slowly, showing that the excluded devices are affected by extended outages rather than by occasional gaps. Since we release the time series for all 70 sensors together with the code implementing the mask, users can reproduce these curves and select the pair $(w, x)$ that best matches the requirements of their application.}

\section{\textcolor{black}{Analyses on POIs confidence}}
\label{app:pois}
\begin{figure}[h]
    \centering
    \includegraphics[width=0.6\linewidth]{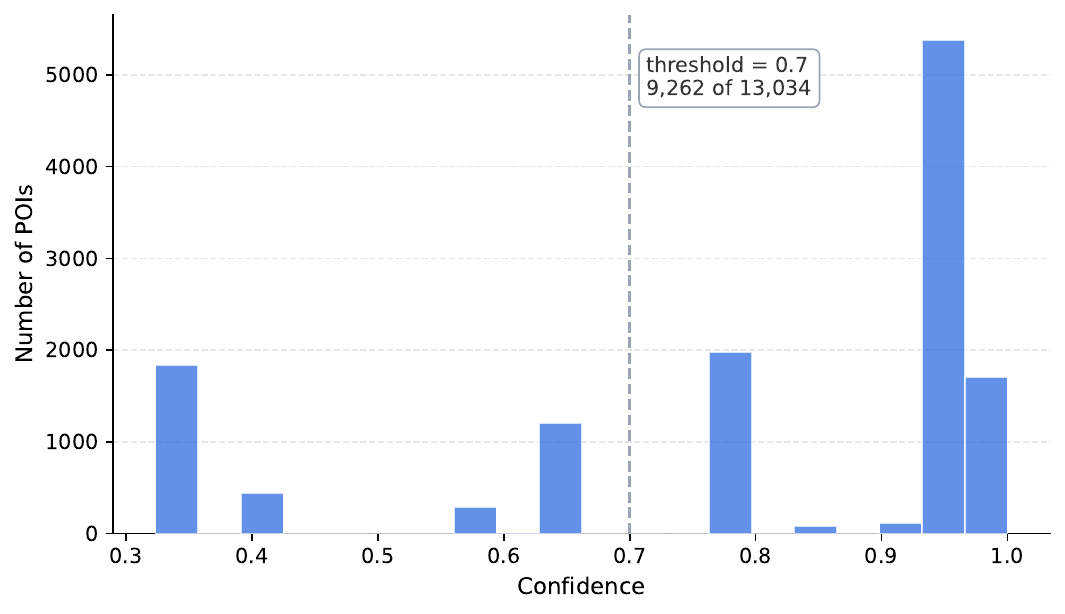}
    \caption{\textcolor{black}{Number of POIs per confidence value.}}
    \label{fig:confidence_histogram}
\end{figure}

\textcolor{black}{Figure \ref{fig:confidence_histogram} reports the number of POIs per confidence value, showing a strongly discretized distribution with two modes: a lower one gathering entries of doubtful existence, and a larger upper mode including POIs whose existence is more certain. The value $0.7$ adopted in Section~\ref{sec:context} as exclusion threshold lies in the middle of a wide plateau extending from roughly $0.65$ to $0.77$, which means that every value in this range leads to the same number of excluded POIs, that is, 3772. The filtering step is therefore insensitive to the precise value of the threshold, and the retained set is dominated by high-confidence entries. We release both raw POIs data and the code implementing the filter, allowing users to use a different threshold for their specific needs.
}  
 
\section{Additional Analyses on Traffic Flow Statistics}
\label{app:flow_analyses}

\textcolor{black}{The statically aggregated statistics defined in Section \ref{sec:flow} induce a weighted directed graph $\mathcal{G} = (\mathcal{V}, \mathcal{E})$ over the traffic sensors, where an edge $(s_i, s_j)$ exists whenever vehicle transitions from $s_i$ to $s_j$ were observed. Throughout this section we denote by $p(s_i,s_j)$, $C(s_i,s_j)$ and $\mu(s_i,s_j)$ the static counterparts of Equations \eqref{prob_matrix}, \eqref{eq:outflow},   and \eqref{eq:travel_times}, i.e., the quantities aggregated over the first 80\% of the observation period rather than within an hourly bin $\tau$.}

\begin{figure}[h!]
    \centering
    \begin{subfigure}[b]{0.46\linewidth}
        \centering
        \includegraphics[width=\linewidth]{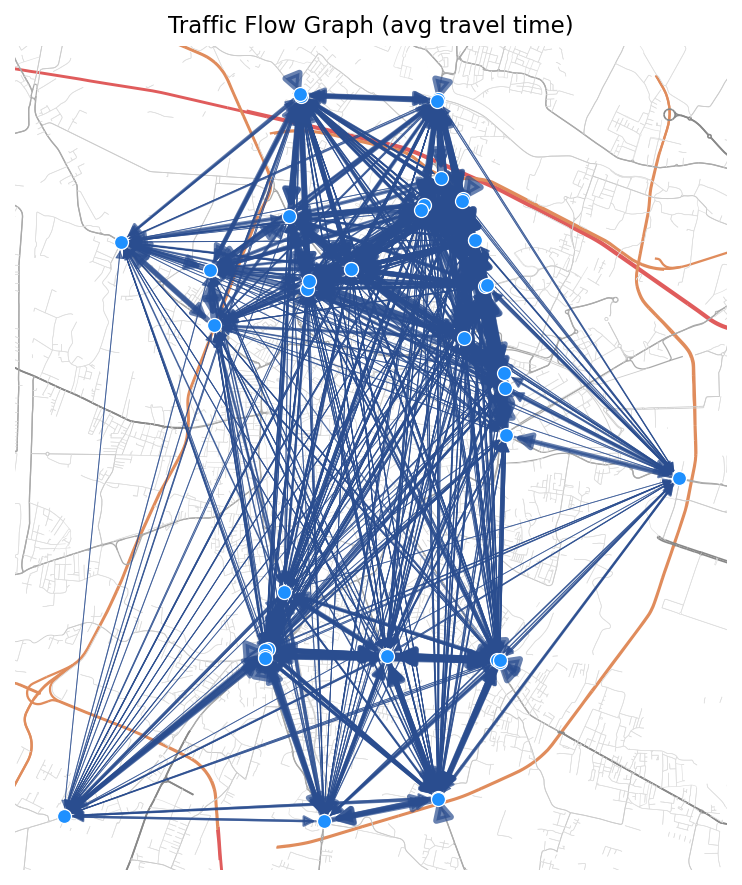}
        \caption{}
        \label{fig:avg_tt_graph}
    \end{subfigure}
    \hfill
    \begin{subfigure}[b]{0.46\linewidth}
        \centering
        \includegraphics[width=\linewidth]{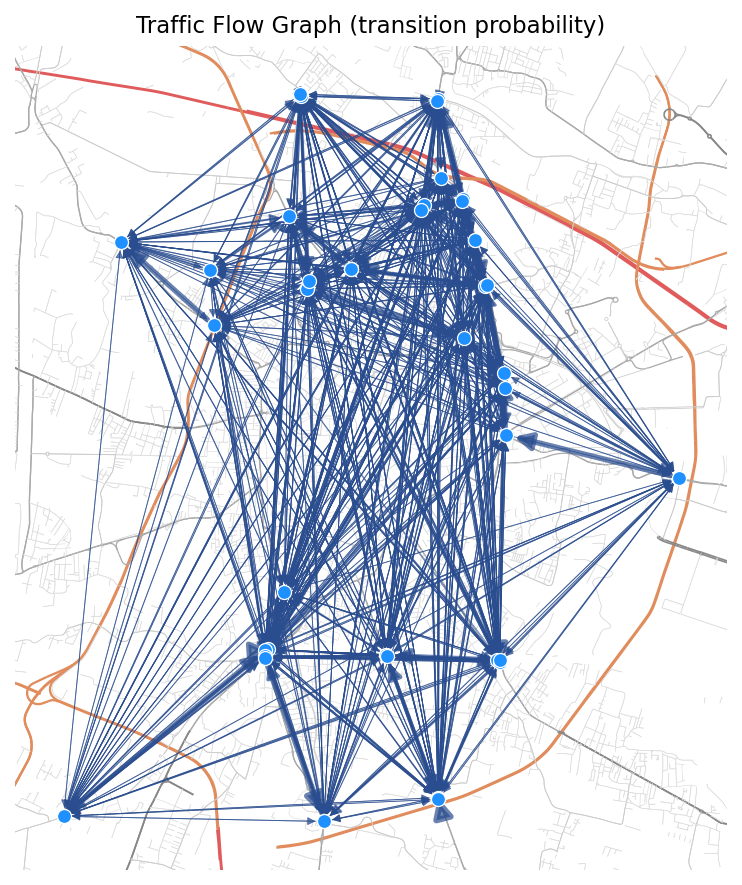}
        \caption{}
        \label{fig:prob_graph}
    \end{subfigure}

    \vspace{0.4cm}

    \begin{subfigure}[b]{0.46\linewidth}
        \centering
        \includegraphics[width=\linewidth]{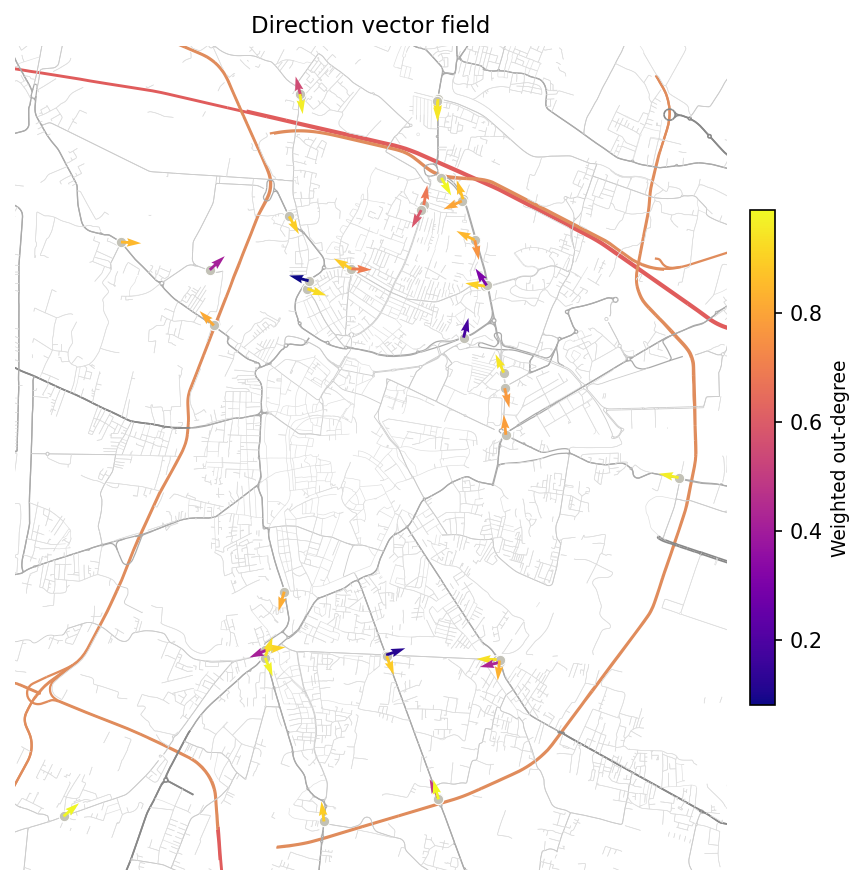}
        \caption{}
        \label{fig:flow_vector_field}
    \end{subfigure}
    \hfill
    \begin{subfigure}[b]{0.50\linewidth}
        \centering
        \includegraphics[width=\linewidth]{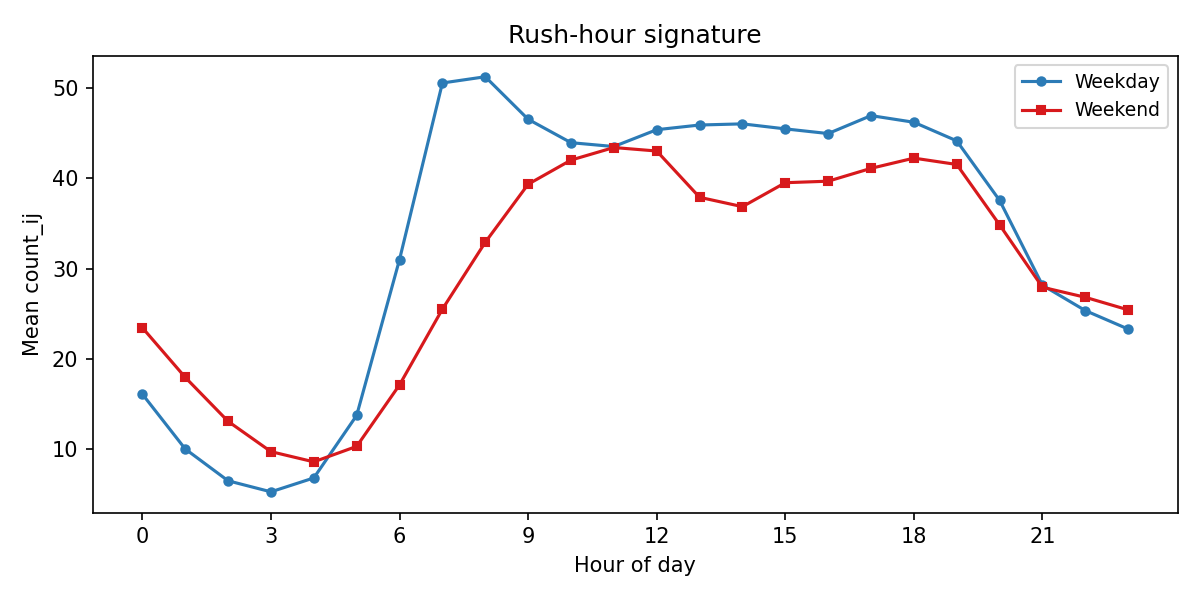}
        \caption{}
        \label{fig:rush_hour_signature}
    \end{subfigure}

    \caption{\textcolor{black}{(a) Flow graph with edges weighted by the Gaussian kernel of Equation \eqref{eq:gaussian_kernel} applied to the average travel times, and (b) the same graph with edges weighted by the transition probabilities; in both panels the arrow width and head size are proportional to the edge weight. (c) Direction vector field obtained from Equation \eqref{eq:vector_field}: arrows are drawn with a fixed length and colored by $\lVert \mathbf{v}_{s_i} \rVert$. Importantly, as $\mathbf{v}_{s_i}$ represents the orientation of the vectors, which gives the prevailing direction of departure from $s_i$, the arrows do not necessarily match the direction of a street. (d) Average number of observed transitions per sensor pair across the hours of the day, on weekdays and on weekend days.}}
    \label{fig:flow_analyses}
\end{figure}

\paragraph{\textcolor{black}{Travel-time graph.}}
\textcolor{black}{To turn average travel times into edge weights we apply a Gaussian kernel, as is standard practice when building graphs over road networks \cite{DBLP:conf/iclr/LiYS018}. Let $\tilde{\mu}(s_i,s_j) = \mu(s_i,s_j) / 3600$ be the average travel time expressed in hours, and let
\[
\sigma = \mathrm{std}\big(\{\tilde{\mu}(s_i,s_j) : (s_i,s_j) \in \mathcal{E}\}\big)
\]
be its standard deviation computed over the edges. The weight of edge $(s_i,s_j)$ is then defined as
\begin{equation}
    w(s_i, s_j) = \exp\!\left( - \left( \frac{\tilde{\mu}(s_i,s_j)}{\sigma} \right)^{\!2} \right),
    \label{eq:gaussian_kernel}
\end{equation}
so that pairs connected by short travel times receive weights close to $1$, while pairs separated by long travel times are progressively down-weighted towards $0$. Figure \ref{fig:avg_tt_graph} draws the resulting graph over the city road network, while Figure \ref{fig:prob_graph} shows the same graph with edges weighted by the transition probabilities $p(s_i,s_j)$ instead.}

\paragraph{\textcolor{black}{Direction vector field.}}
\textcolor{black}{
To summarize the dominant direction of traffic at each sensor, we condense the outgoing edges of every node into a single planar vector. Let $\mathbf{x}_{s_i} \in \mathbb{R}^2$ be the position of sensor $s_i$ in a metric projection of the study area (EPSG:32632), and let $\mathcal{N}_{\text{out}}(s_i) = \{ s_j : (s_i,s_j) \in \mathcal{E} \}$ be its set of downstream neighbors. Denoting by
\[
\mathbf{u}_{ij} = \frac{\mathbf{x}_{s_j} - \mathbf{x}_{s_i}}{\lVert \mathbf{x}_{s_j} - \mathbf{x}_{s_i} \rVert}
\]
the unit vector pointing from $s_i$ towards $s_j$, the flow vector at $s_i$ is given by the transition-probability-weighted sum of the directions of its outgoing edges:
\begin{equation}
    \mathbf{v}_{s_i} = \sum_{s_j \in \mathcal{N}_{\text{out}}(s_i)} p(s_i,s_j)\, \mathbf{u}_{ij}.
    \label{eq:vector_field}
\end{equation}
The orientation of $\mathbf{v}_{s_i}$ gives the prevailing direction of departure from $s_i$, while its magnitude $\lVert \mathbf{v}_{s_i} \rVert$ measures how concentrated that departure is: a node whose outgoing probability mass is concentrated along a single direction yields a long vector, whereas a node redistributing traffic evenly in opposing directions yields a short one. Figure \ref{fig:flow_vector_field} renders the field, drawing every arrow with a fixed length so that the orientations remain legible, and encoding $\lVert \mathbf{v}_{s_i} \rVert$ in the arrow color.}

\paragraph{\textcolor{black}{Rush-hour signature.}}
\textcolor{black}{Finally, Figure \ref{fig:rush_hour_signature} reports the average number of observed transitions per sensor pair, $C_\tau(s_i,s_j)$, aggregated hourly, separately for weekdays and weekend days. Consistently with the traffic volume profiles of Figure \ref{fig:weekdays_weekends}, the weekday curve shows a pronounced morning peak and a broader late-afternoon peak, whereas the weekend curve is markedly flatter and lacks the morning peak entirely. This shows that the rush-hour regularities observed in the volume time series are also recovered by the trajectory-based transition statistics, which are derived through an entirely independent processing pipeline.}

\end{document}